\documentclass[12pt,amssymb]{article} 

\usepackage{mathrsfs}
\usepackage{amssymb}
\usepackage[dvips]{graphicx}

\newcommand{\newsection}[1]{
\addtocounter{section}{1} \setcounter{equation}{0}
\setcounter{subsection}{0} \addcontentsline{toc}{section}{\protect
\numberline{\arabic{section}}{{\rm #1}}} \vglue .6cm \pagebreak[3]
\noindent{ \bf  \thesection. #1}\nopagebreak[4]\par\vskip .3cm}
\newcommand{\newsubsection}[1]{
\addtocounter{subsection}{1}\setcounter{subsubsection}{0}
\addcontentsline{toc}{subsection}{\protect
\numberline{\arabic{section}.\arabic{subsection}}{#1}} \vglue .4cm
\pagebreak[3] \noindent{\it \thesubsection.
#1}\nopagebreak[4]\par\vskip .3cm}

\newcommand{\newsubsubsection}[1]{
\addtocounter{subsubsection}{1}
\addcontentsline{toc}{subsubsection}{\protect
\numberline{\arabic{section}.\arabic{subsection}.\arabic{subsubsection}}{
#1}} \vglue .4cm \pagebreak[3] \noindent{\it \thesubsubsection.
#1}\nopagebreak[4]\par\vskip .3cm}

\makeatletter
\newcommand{\seclabel}[1]{%
  \@bsphack
  \protected@write\@auxout{}%
     {\string\newlabel{#1}{{\thesection}{\thepage}}}
  \@esphack
  }
\newcommand{\subseclabel}[1]{%
  \@bsphack
  \protected@write\@auxout{}%
     {\string\newlabel{#1}{{\thesubsection}{\thepage}}}
  \@esphack
  }
\newcommand{\tablabel}[1]{%
  \@bsphack
  \protected@write\@auxout{}%
     {\string\newlabel{#1}{{\arabic{tabnum}}{\thepage}}}
  \@esphack
  }
\makeatother

\renewcommand{\theequation}{\thesection.\arabic{equation}}

\newlength{\extraspace}
\newlength{\extraspaces}
\newcounter{dummy}
\newcommand{\bc}{\begin{center}}
\newcommand{\ec}{\end{center}}
\newcommand{\be}{\begin{equation}
\addtolength{\abovedisplayskip}{\extraspaces}
\addtolength{\belowdisplayskip}{\extraspaces}
\addtolength{\abovedisplayshortskip}{\extraspace}
\addtolength{\belowdisplayshortskip}{\extraspace}}
\newcommand{\ee}{\end{equation}}

\newcommand{\ba}{\begin{eqnarray}
\addtolength{\abovedisplayskip}{\extraspaces}
\addtolength{\belowdisplayskip}{\extraspaces}
\addtolength{\abovedisplayshortskip}{\extraspace}
\addtolength{\belowdisplayshortskip}{\extraspace}}
\newcommand{\ea}{\end{eqnarray}}

\newcommand{\is}{& \!\! = \!\! &}
\newcommand{\ban}{\begin{eqnarray*}
\addtolength{\abovedisplayskip}{\extraspaces}
\addtolength{\belowdisplayskip}{\extraspaces}
\addtolength{\abovedisplayshortskip}{\extraspace}
\addtolength{\belowdisplayshortskip}{\extraspace}}
\newcommand{\ean}{\end{eqnarray*}}
\newcommand{\baa}{
\addtocounter{equation}{1} \setcounter{dummy}{\value{equation}}
\setcounter{equation}{0}
\renewcommand{\theequation}{\thesection.\arabic{dummy}\alph{equation}}
\begin{eqnarray}
\addtolength{\abovedisplayskip}{\extraspaces}
\addtolength{\belowdisplayskip}{\extraspaces}
\addtolength{\abovedisplayshortskip}{\extraspace}
\addtolength{\belowdisplayshortskip}{\extraspace}}
\newcommand{\eaa}{
\end{eqnarray}
\setcounter{equation}{\value{dummy}}
\renewcommand{\theequation}{\thesection.\arabic{equation}}}

\input epsf

\newcounter{fignum}
\newcounter{tabel}

\newcounter{tabnum}
\newcommand{\vev}[1]{\left\langle #1\right\rangle}

\newcommand{\half}{\frac{1}{2}}
\newcommand{\del}{\partial}
\newcommand{\delb}{\bar{\del}}
\newcommand{\eol}{\nonumber \\}

\newcommand{\cO}{{\cal O}}
\newcommand{\Ext}{{\rm Ext}}
\newcommand{\Hom}{{\rm Hom}}

\newcommand{\bt}{{\bf 10}}
\newcommand{\bfv}{{\bf 5}}
\newcommand{\bfb}{{\overline{\bf 5 \!}\,}}
\newcommand{\btb}{{\overline{\bf 10 \!}\,}}

\begin{document}

%
%

\begin{flushright}
December 2011\\
\end{flushright}
\vspace{1.5cm}

\thispagestyle{empty}

%
%

\begin{center}
{\Large\bf  Gluing Branes II: \\
Flavour Physics and String Duality
 \\[13mm] }

{\sc Ron Donagi}\\[2.5mm]
{\it Department of Mathematics, University of Pennsylvania \\
Philadelphia, PA 19104-6395, USA}\\[9mm]

{\sc Martijn Wijnholt}\\[2.5mm]
{\it Arnold Sommerfeld Center, Ludwig-Maximilians Universit\"at\\
Theresienstrasse 37 \\
D-80333 M\"unchen, Germany }\\
[20mm]

 {\sc Abstract}

\end{center}

Recently we discussed new aspects of degenerate brane
configurations, which can appear in the context of heterotic
strings, perturbative type II, or $M$/$F$-theory. Here we continue
our study of degenerate brane configurations, focussing on two
applications. First we show how the notion of gluing can be viewed
as a tool to engineer flavour structures in $F$-theory and type
IIb, such as models with bulk matter and with Yukawa textures
arising from the holomorphic zero mechanism. We find that there is
in principle enough structure to solve some of the major flavour
problems without generating exotics. In particular, we show how
this addresses the $\mu$-problem, doublet/triplet splitting and
proton decay. Secondly, we describe the Fourier-Mukai transform of
heterotic monad constructions, which occur in the large volume
limit of heterotic linear sigma model vacua. Degenerate structures
again often appear. One may use this to explore strong coupling
phenomena using heterotic/$F$-theory duality.

\vfill

\newpage

\renewcommand{\Large}{\normalsize}

\tableofcontents

\newpage

\newsection{Introduction}

The present paper is a continuation of our study of degenerate
brane configurations. In \cite{Donagi:2010pd} we pointed out the
important role of the gluing morphism. In reference
\cite{DegBranesI} we considered theoretical aspects more
systematically, including and generalizing configurations studied
in earlier work such as
\cite{DEL,Aspinwall:1998he,Donagi:2003hh,Cecotti:2010bp,Chiou:2011js,
Bershadsky:1997ec,Bershadsky:1997zs,Berglund:1998ej,Aspinwall:2000kf}.
Amongst others, in reference \cite{DegBranesI} we discussed exact
sequences associated to gluing operations, aspects of stability and
the hermitian Yang-Mills-Higgs metric, and walls of marginal
stability. In the present paper we focus on certain applications.

There were in fact two independent lines of inquiry which motivated
us to take a closer look at degenerate configurations. The first
motivation is to get a more systematic understanding of flavour
structure in $F$-theory, or more generally models with intersecting
branes. The second is to get a better understanding of certain string
dualities. It will hopefully become clear that degenerate structures
are ubiquitous and critical for
the phenomenology of string compactifications. With the tools
developed in part I, we have at least in principle everything we need to analyze them.

\newsubsection{Flavour structure and degeneration}

One of the main motivations for top-down models is the possibility
to get some understanding of the origin of flavour structure. In
particular, there are various hints that flavour should be
generated near the GUT scale. So it is a natural question to ask
what kind of flavour structures can naturally occur in Kaluza-Klein
GUT models. In such models, the higher dimensional theory is rather
constrained (in order for it to have a known UV completion), and
much of the four-dimensional physics can be traced back to the
geometry of the compactification. Flavour structure gets related to
the geometric properties of wave functions in the extra dimensions.

Let us consider this question in $F$-theory. Here one grows four
extra dimensions at the GUT scale. Thus we focus on an
eight-dimensional gauge theory which is compactified on a
four-manifold $S$ down to four dimensions. Supersymmetric
configurations are given by $K_S$-valued Higgs bundles on a complex
manifold $S$. In generic $SU(5)_{GUT}$ models, the gauge fields
propagate in the bulk of $S$ and the $\bt$ and $\bfb$ matter fields
are confined to Riemann surfaces on $S$, called the matter curves.

Using a general equivalence between supersymmetric ALE fibrations,
Higgs bundles and spectral covers
\cite{Donagi:2009ra,Pantev:2009de}, such local models may be
represented by a configuration of holomorphic 7-branes in an
auxiliary Calabi-Yau three-fold, or as an ALE-fibration over $S$
with flux. For $SU(5)$ GUT models this ALE fibration is generically
of the form:
\be\label{SU5GUTALE} y^2 \ = \ x^3 + b_0 z^5 + b_2 z^3 x + b_3 z^2 y
+ b_4 z x^2 + b_5 xy \ee
where the $b_i$ are complex polynomials on $S$. Matter in the $\bt$ or $\btb$ is
confined to the Riemann surface given by $b_5 = 0$, and matter in the
$\bfb$ or $\bfv$ is confined to $b_0 b_5^2 -b_2 b_3 b_5 + b_3^2 b_4 = 0$.
We have explained elsewhere how to embed such models in a global compactification
\cite{Donagi:2009ra}, effectively by using Tate's algorithm \cite{Tate,Bershadsky:1996nh} in
reverse. However, in this paper we will be interested in aspects of flavour which
must already be present in the local model. With suitable Noether-Lefschetz fluxes,
the moduli appearing in the $b_i$ are stabilized at isolated critical points of the
superpotential. We have estimated that one can construct
at least $10^{1000}$ models of this form with the spectrum of the MSSM.

Now a priori one might have thought that for generic values of the
moduli there is no flavour structure whatsoever. It appears that the
situation is actually better than that. The main point is that the
degree (and hence the volume) of the matter curve for the $\bfv$ and
$\bfb$ is larger than the degree of the matter curve of the $\bt$.
Then one should expect that upon proper normalization of the kinetic
terms, the $\bt \cdot \bfb\cdot \bfb$ down Yukawas are slightly
suppressed compared to the $\bt \cdot \bt \cdot \bfv$ up type
Yukawas \cite{Hayashi:2009bt}, by a factor
\be
\lambda_d/\lambda_u \ \sim\ \sqrt{\deg_{\bt}/\deg_{\bfb}}
\ee
This goes clearly in the right
direction and is much better than the situation in type IIb. Type
IIb GUT models actually also exhibit a flavour structure, but it
predicts that the top quark Yukawa coupling is exponentially
suppressed with respect to the down coupling in the natural
expansion parameter, the string coupling constant.

In any case, even if the above idea is correct, the hierarchy is not
parametric (at least if we keep the degrees of the $b_i$ fixed),
and such a generic model does not seem to be able to
naturally explain any additional flavour structures. It is clear
that we need some extra structure, and quite a number of ideas have
been explored in the literature.

The basic idea for getting extra structure is to degenerate the generic
models in some way. In $F$-theory, the most obvious ingredients we can degenerate
are the flavour branes, by varying the $b_i$. For instance for certain degenerations of
the flavour branes one may get an extra light $U(1)$ symmetry
\cite{Tatar:2006dc}, which imposes selection rules on the Yukawa
couplings. For other degenerations of the flavour branes, one may
get matter in the bulk of a 7-brane, instead of on the intersection
of two 7-branes, and again this implies extra flavour structure.
Models with bulk matter were previously studied in
\cite{Donagi:2008ca,Beasley:2008dc}, but the results were not too
encouraging.

Now when one considers degenerate configurations, one must be
careful to include all the ingredients, as the rules are a little
less obvious than for generic configurations. Indeed as found in
\cite{Donagi:2010pd}, even the simplest possible flavour structures were not
correctly understood: it turns out that the gluing data was missed,
even though this appears very naturally in degenerations of more
generic models.

The first half of the present paper focuses on flavour structures,
using the improved understanding of degenerate configurations
developed in part I. The main goal of these sections is to show
that there is in principle enough structure to solve the major
flavour problems, without generating exotics and destroying
unification.

In particular, our improved understanding shows how to implement
the holomorphic zero mechanism in intersecting brane
configurations. The idea of using such holomorphic zeroes in
$F$-theory (or brane configurations) was already discussed in
\cite{Tatar:2006if,Kuriyama:2008pv}, but we believe the idea was
not used to its full extent, and a few aspects had not been clear.
In section \ref{FlavModel} we will further show how one can
simultaneously address the problems of $R$-parity, dimension five
proton decay, the $\mu$-problem, doublet/triplet splitting and a
simple flavour hierarchy in $F$-theoretic GUTs (or the
corresponding heterotic models). In particular, we will engineer a
superpotential of the form
\be
W \ = \ \bt_m \, \bt_m \, \bfv_h\ +\ {\vev{X}\over M} \bt_m \, \bfb_m \, \bfb_h\
+\ {\vev{X}\over M^2} \bt_m\, \bt_m \, \bt_m \, \bfb_m
\ee
in which no $R$-parity violating terms appear. $R$-parity however
is not preserved in the K\"ahler potential. Thus such models
ultimately do predict some form of $R$-parity violation, which
could have very interesting phenomenological consequences.

\newsubsection{Strong coupling phenomena in $F$-theory}

$F$-theory is only understood perturbatively as a large volume
expansion. Clearly it would be of interest to get a better
non-perturbative understanding, and explore the theory in other
corners of the K\"ahler moduli space. The best available tool for
this is heterotic/$F$-theory duality. Comparing BPS states yields
the identification
\be \lambda_8 \ = \ V_{{\bf P}^1} \ee
where $\lambda_8$ is the eight-dimensional heterotic string coupling
and $V_{{\bf P}^1}$ is the volume of the base of the elliptically
fibered $K3$ on the $F$-theory side, measured in Planck units.
$F$-theory is weakly coupled in the limit of large $V_{{\bf P}^1}$,
and the heterotic string is weakly coupled in the limit of small
$V_{{\bf P}^1}$.

One important technique for constructing heterotic vacua is the
linear sigma model, which yields monad constructions in the
geometric regime. In such models, the $(0,2)$ CFT is relatively
well-understood and can be extrapolated to corners of the K\"ahler
moduli space where curvatures are large. Such models also exhibit a
number of interesting dualities. Now in order to map this to an
$F$-theory model we need the associated spectral cover, and it turns
out that the spectral cover for heterotic monad constructions is
often degenerate \cite{Bershadsky:1997zv}. (Our investigations
actually indicate this is not the general situation, and the
examples that were worked out were just too special).

The apparent conclusion in the nineties was that the $F$-theory
duals would be sick in some way. We would like to emphasize that
this is not the case, although it is true that the $11d$ supergravity
description of $F$-theory can be problematic. As explained in \cite{DegBranesI}, a degenerate cover gives rise to a
smooth $8d$ non-abelian gauge theory
configuration provided a suitable stability condition is satisfied, and therefore
many configurations with degenerate spectral covers make perfect sense in $F$-theory.
However to understand the $F$-theory dual it is crucial that we obtain the
spectral sheaf rather than the spectral cover, and the spectral
sheaf for monad bundles has hitherto not been understood.

This gives us the second reason to revisit degenerate covers.
Interesting enough, the degenerations that we study in the context
of flavour also show up quite naturally as the Fourier-Mukai
transform of standard heterotic constructions, such as the standard embedding
and the construction of bundles by extension, even though they correspond to smooth
solutions of the hermitian Yang-Mills equations. The fact that the standard embedding
gives rise to such structures seems to us so fundamental, that we will spend a whole section
examining a particular example. Recall also from part I that degenerate
structures play an important role in understanding walls of marginal stability.
Altogether this serves to
illustrate that many degenerate configurations are not only perfectly acceptable, they are in fact
a critical aspect of the phenomenology of string compactifications, because they tend
to imply additional structure.

It is clear that there is a lot of interesting work to be done
comparing heterotic linear sigma model vacua with $F$-theory. In
order to finish this paper in a finite amount of time however, we
will only focus on establishing the technology. To this end, we will
give a prescription for deriving the complete spectral data of monad
bundles, extending previous work of
\cite{Bershadsky:1997zv,Lazaroiu:1997}.

\newpage

\newsection{Bulk matter revisited}

\seclabel{BulkMatter}

\newsubsection{$SO(10)$-models}

\subseclabel{SO(10)Models}

One of the main motivations for this project was to reconsider the
issue of chiral matter in the bulk of a 7-brane in light of the
gluing morphism. Bulk matter in $F$-theory models was originally
discussed in \cite{Donagi:2008ca,Beasley:2008dc}.

Apart from the interest in bulk matter and flavour structure,
according to the analysis in \cite{Donagi:2009ra} it is also much
easier to decouple gravity and hidden sectors if additional sheets
of the spectral cover coincide with $S$. Even for $SO(10)$ models
the constraints coming from the GUT divisor being contractible are
already much weaker than for $SU(5)$ models.

However if the gluing morphisms vanish, it turns out to be difficult
to avoid exotics, and the constraints on the interactions can be too
stringent. To exemplify this, in the first part of this section we
would like to focus on $SO(10)$ models with the breaking of the GUT
group done by fluxes. For such models, it was argued by Beasley,
Heckman and Vafa \cite{Beasley:2008kw} that there are always exotics
in the bulk. In retrospect, the argument assumed that the gluing
morphism on $C_1 \cap C_4$ vanishes.

In spite of the length of this section, our main point is very
simple: once the gluing morphism is non-zero, modes in the bulk and
on the matter curves are not independent. In the special class of
examples that we will consider, where the gluing morphism can be
turned off continuously, one can see explicitly how the bulk exotics
can (and generically will) pair up with modes that are localized on
a matter curve as we turn on the gluing. In the following
discussion, we explain how this works in detail, and address some
additional issues arising for degenerate configurations along the
way.

\newsubsubsection{Review of the problem of exotics}

Let us therefore reexamine $SO(10)$ models. The terminology is
perhaps slightly misleading, because we never have an unbroken $4d$
$SO(10)$ group. What we mean is that the $F$-theory geometry has an
$SO(10)$ singularity along the GUT cycle, with the remaining
breaking due to non-trivial $G$-flux. In terms of the $E_8$ gauge
theory, this means that the spectral cover $C_5$ for the $SL(5,{\bf
C})$ Higgs bundle which breaks $E_8$ to $SU(5)_{GUT}$ is reducible,
i.e. we have $C_5 = C_4 \cup C_1$ where $C_1 = S$ and $C_4$ is a
non-trivial spectral cover for $SU(4) \subset E_8$. Such models can
be analyzed in two steps, first breaking $E_8$ to $SO(10)$, where
we use the decomposition
\be {\bf 248} \ = \ ({\bf 1},{\bf 45}) + ({\bf 15},{\bf 1}) + ({\bf
4},{\bf 16}) + (\overline{\bf 4\!}\,, \overline{\bf 16\!}\,) + ({\bf
6},{\bf 10}) \ee
Even though the spectral cover is reducible, the spectral sheaf and
hence the $SL(5,{\bf C})$ Higgs bundle need not be reducible. The
remaining breaking to $SU(5)$ and further to $SU(3)\times
SU(2)\times U(1)$ is done by the spectral sheaf.

\begin{figure}[t]
\begin{center}
            \scalebox{.45}{
               \includegraphics[width=\textwidth]{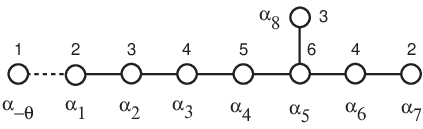}
               }
\end{center}
\vspace{-.5cm} \caption{ \it The extended $E_8$ Dynkin diagram and
Dynkin indices.}\label{E8Dynkin}
\end{figure}

Ignoring the fluxes, such a configuration leaves two unbroken
$U(1)$'s that commute with the Standard Model gauge group. We label
the coroots of $E_8$ by $\omega_i$, where $\omega_i(\alpha_j) =
\delta_{ij}$. The structure group of our spectral cover is
$W_{A_3}$, the Weyl group  generated by the Weyl reflections
associated to $\{\alpha_{-\theta}, \alpha_1,\alpha_2 \}$, see figure
\ref{E8Dynkin}. Written in terms of the $\omega_i$, we can take the
unbroken $U(1)$'s to be
\be \omega_Y\ =\ \omega_4 - {5\over 6} \omega_5, \qquad \omega_{B-L}
\ =\ \omega_3 - {4\over 5} \omega_4 \ee
We normalized $B-L$ so that the right-handed neutrino has charge
one. The matter curves of a generic $SU(5)$ model split up in the
following way:
\be
\begin{array}{rclcl}
  S\cdot C_5 & = & S \cdot S + \Sigma_{\bf 16} & =& -c_1 + (2c_2-t) = c_1-t
  \\[2mm]
  S \cdot C_{10} & = & \Sigma_{\bt_v} + \Sigma_{\bf 16} & =& (6c_1-2t) +
(2c_1-t) = 8c_1 - 3t
\end{array}
\ee
where $c_1 = c_1(T_S)$ and $t= c_1(N_S)$. Thus the $\bt$ will split
up into a $\bt_{-4/5}$ localized in the bulk and a $\bt_{1/5}$
localized on $\Sigma_{\bf 16}$. Similarly the $\bfb$ splits up into
a $\bfb_{2/5}$ localized on $\Sigma_{\bt_v}$ and a $\bfb_{-3/5}$
localized on $\Sigma_{\bf 16}$. This of course fits neatly into
$SO(10)$ representations, as follows:
\be
\begin{array}{rcl}
  \Sigma_{\bf 16}: &  & {\bf 16}\ =\ {\bf 1}_{1} + \bt_{1/5} + \bfb_{-3/5}
  \\[2mm]
  \Sigma_{\bt_v}: &  & {\bf 10}\ =\ \bfb_{2/5} + \bfv_{-2/5} \\[2mm]
  {\rm bulk}: & & {\bf 45}\ =\ {\bf 24}_0 + {\bf 1}_0 + \bt_{-4/5} + \btb_{4/5}
\end{array}
\ee
We will denote the singlets with charge one by $N$ and the singlets
with charge minus one by $\bar N$.

With these decompositions, we can look for exotic matter. We denote
the restriction of the $U(1)_Y$ line bundle to $S$ by $L_Y$, and the
restriction of the $U(1)_{B-L}$ line bundle to $S$ by $L_{B-L}$. The
first part of the argument is the same as for $SU(5)_{GUT}$ models
with hypercharge flux. On a Del Pezzo surface $dP$, for any line
bundle $M$ we have
\be \chi(dP, M)\ =\ 1 + \half c_1(M)^2 - \half c_1(M) \cdot c_1(K)
\ee
We consider the off-diagonal modes $({\bf 3},{\bf 2})_{5/6,0}$ from
breaking the adjoint of $SU(5)_{GUT}$ to $SU(3)_c \times SU(2)_w
\times U(1)_Y$. These modes live purely in the bulk of $S$, so
$L_Y^{5/6}$ is integer quantized. The number of $({\bf 3},{\bf
2})_{5/6,0}$ modes is counted by $-\chi(dP,L_Y^{5/6})$, and absence
of the $({\bf 3},{\bf 2})_{5/6,0}$ modes and their conjugates then
requires that
\be \chi(dP,L_Y^{5/6}) + \chi(dP,L_Y^{-5/6})\ =\ 2\left(1 + \half
c_1(L_Y^{5/6})^2\right)\ =\ 0. \ee
This implies the restrictions $c_1(L_Y^{5/6}) \cdot c_1(K_S) = 0$
and $c_1(L_Y^{5/6})^2 = -2$, well-known from hypercharge flux
breaking of $SU(5)_{GUT}$ models
\cite{Beasley:2008kw,Donagi:2008kj}.

Now in addition consider the bulk modes coming from the
$\btb_{4/5}$. Under $SU(3)_c \times SU(2)_w \times U(1)_Y \times
U(1)_{B-L}$ they split up as
\be \btb_{4/5}\ =\ (\overline{\bf 3 \!}\,, {\bf 2})_{-1/6,4/5} +
({\bf 1}, {\bf 1})_{-1,4/5} + ({\bf 3}, {\bf 1})_{2/3,4/5} \ee
It follows that the first Chern class of $Q \equiv L_Y^{1/6} \otimes
L_{B-L}^{-4/5}$ must also be integer quantized. Furthermore, absence
of states in the $\btb_{4/5}$ requires that
\be\label{SO(10)BulkExotics}
\begin{array}{lcc}
  \chi({dP},Q^{-1}\otimes L_Y^{-5/6}) & = & 0 \\
  \chi({dP},Q^{-1}) & = & 0 \\
  \chi({dP},Q^{-1} \otimes L_Y^{5/6})  & = & 0
\end{array}
\ee
By considering the linear combination (line $1$ + line $3\, -\,
2\times$ line $2$), we find that $c_1(L_Y^{5/6})^2=0$, a
contradiction \cite{Beasley:2008kw}. The argument can even be
extended to slightly more general configurations, because the same
algebra works when $Q^{-1}$ is a higher rank bundle.

\newsubsubsection{Turning on the gluing morphism}

The above argument implicitly assumes the vanishing of the gluing
morphism, here the VEV of the field $\bar N$. When this VEV is
non-zero, it is no longer true that the bulk modes are counted by
(\ref{SO(10)BulkExotics}). However the zero modes are still counted
by infinitesimal deformations of the Higgs bundle, and for the case
at hand this specializes to the usual Ext groups known from generic
models \cite{Donagi:2009ra}. The only difference is that the support
of the spectral sheaf has degenerated, and we need to take some more
care in the computation.

As we discussed in section 2.7 of part I, generally the gluing
morphism is only a meromorphic section of $L_1^\vee \otimes
L_2|_\Sigma$, and in this case we need to use a sequence of the form
\be\label{RedFundShortSeq} 0\ \to\ {\cal L}\ \to\ i_{1*}L_1 \oplus
i_{2*}L_2\ \to\ i_{\Sigma *}L_\Sigma\ \to\ 0 \ee
to find the
spectrum. Let us assume here that it is given by a holomorphic
section, so that we can use the extension sequence
\be\label{RedShortSeq} 0 \ \to \ i_{2*}L_2(-\Sigma) \ \to\ {\cal L}
\ \to \ i_{1*}L_1 \ \to \ 0 \ee
instead. Although
not the general case, it is already sufficient to show that the
exotics can get lifted, and simplifies the calculations.

Then the correct way to count the zero modes proceeds as in section
2.7 of part I. As the gluing VEV is holomorphic, we may turn it
off (while ignoring $D$-flatness), compute the modes in the bulk and
on the matter curve, and then turn the gluing VEV back on. Then
there are effectively two changes in the computation of exotics.
First, the bulk spectrum is not necessarily computed by
(\ref{SO(10)BulkExotics}), because $D$-flatness is not satisfied
when we turn off the gluing VEV. Secondly, candidate bulk modes may
be lifted by a superpotential coupling between the modes in the bulk
and the modes on the matter curve:
\be \btb_{4/5}\, \bt_{1/5}\, \bar N_{-1} \ee
when we turn on a VEV for the gluing morphism $\bar N$. (More
precisely, we should decompose this coupling under the Standard
Model gauge group, because the $SU(5)_{GUT}$ is broken explicitly by
the fluxes. We will see this in more detail below).

Although plausible, one should not assume that the above superpotential coupling is automatically
present. Indeed we will see examples where certain couplings are generically absent despite
the fact that they are allowed by the gauge symmetries, and we will make good use of that to
solve the mu-problem. Thus one of the
main things to check below is that the above superpotential coupling is
indeed present in the brane configuration.

 \begin{figure}[th]
\begin{center}
            \scalebox{.5}{
               \includegraphics[width=\textwidth]{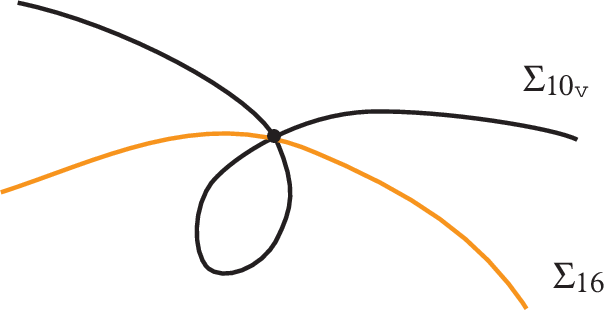}
               }
\end{center}
\vspace{-.5cm}
\begin{center}
\parbox{14cm}{\caption{ \it The curve on which the $\bfb$ or $\bfv$ matter of an $SU(5)$
GUT propagates has factorized into two pieces, but the modes on
these two curves are not independent if the gluing morphism is
non-zero. Similarly, modes in the $\bt$ or $\btb$ of $SU(5)$ seem to
originate from the bulk or from a matter curve, but are not
independent when the gluing morphism is
non-zero.}\label{SO10Curves}}
\end{center}
 \end{figure}
Although our main interest here is in the $\bt$ and $\btb$, there is
a similar story for the $\bfv$ and $\bfb$ fields. Ignoring the
hypercharge flux to simplify the discussion, these modes are
computed by
\be \Ext^p(\cO_S, i_{10*}L_{C_{10}}) \ee
as usual. The sheaf $L_{C_{10}}$ is supported on a ten-fold covering
$C_{10}$. The only difference with the generic case is that $C_{10}$
happens to be reducible, and so we should apply the discussion in
section 2.7 of part I 
to compute these zero modes. The cover
$C_{10}$  splits as a six-fold covering $C_6$, and a
fourfold-covering $C_4$ which we have already met above. Thus the
$\bfv$ and $\bfb$ fields are supported on two seemingly different
matter curves, the $\bfb_{2/5}$ on $S \cap C_6=\Sigma_{\bt_v}$, and
the $\bfb_{-3/5}$ on $S\cap C_4 =\Sigma_{\bf 16}$. However with a
non-vanishing gluing VEV, the zero modes on these two curves are not
independent, and must be glued along the intersection
$\Sigma_{\bt_v} \cap \Sigma_{\bf 16}$. Let us spell this out in some
detail. In our case there is the additional complication that the
curve $\Sigma_{\bt_v}$ has a double point singularity at the
intersection, and there is some intricate group theory involved.
This is a known behaviour which is dealt with by
lifting to the normalization. In the following we assume a simple
intersection between smooth curves.

Then, abstractly we have a sheaf $N$ on $\Sigma = \Sigma_1 \cup
\Sigma_2$, restricting to $N_1$ and $N_2$ respectively, and we are
interested in
\be H^0(\Sigma, N) \ee
Using the long exact sequence, we find
\be\label{Seq1}
\begin{array}{ccccccccc}
  0 & \to & H^0(\Sigma_2, N_2(-p)) & \to & H^0(\Sigma, N) & \to & H^0(\Sigma_1, N_1)  & &
  \\[3mm]
   & \to  &H^1(\Sigma_2, N_2(-p))  & \to & H^1(\Sigma, N) & \to
   & H^1(\Sigma_1, N_1)  & \to & 0
\end{array}
\ee
The intersection $\Sigma_1 \cap \Sigma_2$ is a number of points, and
the gluing morphism (restricted to $\Sigma$) is a complex number for
each intersection point $p$. For simplicity we assume there is only
a single intersection point, the generalization being obvious. We
also have a second long exact sequence on $\Sigma_2$:
\be\label{Seq2}
\begin{array}{ccccccccc}
  0 & \to & H^0(\Sigma_2, N_2(-p)) & \to & H^0(\Sigma_2, N_2) & \to & H^0(p, N_2|_p)  & &
  \\[3mm]
   & \to  &H^1(\Sigma_2, N_2(-p))  & \to & H^1(\Sigma_2, N_2) & \to
   & 0 & &
\end{array}
\ee
Now the coboundary map in (\ref{Seq1}) representing the Yukawa
coupling is given as follows. We take a section $s_1 \in
H^0(\Sigma_1, N_1)$, restrict it to $p$, multiply by the gluing
morphism (a complex number) to get a generator in $H^0(p,N_2|_p)$,
and then compose with the coboundary map in (\ref{Seq2}). Then we
see that if this map is zero, so that $s_1$ is not lifted, then it
defines a generator in $H^0(p,N_2|_p)$ which can be extended to a
section $s_2 \in H^0(\Sigma_2, N_2)$, the extension is unique up to
sections which vanish at $p$, and $s_1$ and $s_2$ agree at $p$ up to
multiplication by the gluing morphism. From (\ref{Seq1}) we further
see that the remaining generators of $H^0(\Sigma,N)$ are sections
$s_2$ of $N_2$ that vanish at $p$. Thus we derived from first
principles the expected statement: generators of $H^0(\Sigma,N)$ are
given by pairs of sections $(s_1,s_2)$ on $\Sigma_1$ and $\Sigma_2$
separately, which agree at the intersection $p$ up to a complex
number (the value of the gluing morphism). Thus the gluing condition
will normally eliminate some candidate $\bfv$ and $\bfb$ zero modes,
and allows the zero modes to spread over the union of
$\Sigma_{\bt_v}$ and  $\Sigma_{\bf 16}$ if they do not vanish at the
intersection.

The resulting wave functions for zero modes look a bit singular.
This is an artefact of the purely holomorphic description that we
used to construct them. When the Higgs bundle is stable, the actual
solution of Hitchin's equations and the wave functions for the first
order deformations should be smooth.

\newsubsubsection{Building examples}

In order to construct explicit examples, there are basically two
ways to proceed. We could either degenerate a flat family of smooth
models, or we could start with a reducible model and try to turn on
the gluing VEV. Let us focus on the latter.

Let us list the data that we have to specify. We take the base to be a del Pezzo surface.
On it, we start with a reducible configuration
where the VEV of $\bar N$ vanishes. Then the spectral cover really has six sheets:
we have $C_4$ for breaking $E_8\to SO(10)$, $C_1=S$ associated to $U(1)_{B-L}$, and
another copy of $S$ (let's call it $C_0$) to accommodate non-zero $L_Y$. By abuse of notation, we write
$C_6 = C_0 \cup C_1 \cup C_4$ and $C_0\cup C_1=2S$ even though that is not quite correct scheme theoretically.
We need to specify
$C_4$ and a line bundle $L_4$ on it. On $2S= C_0 \cup C_1$ we need to specify a sum of two line bundles:
\be L_Y^{-5/6} \oplus (L_Y^{1/6} \otimes L_{B-L}^{-4/5})\
\equiv \ L_Y^{-5/6} \oplus Q \ee
The field $\bar N$ corresponds to a gluing morphism pointing from $C_4$ to $C_1$.
Turning it on corresponds to forming a new sheaf $L_5$ which is the extension of $i_{4*}L_4$
by $i_{1*}Q$, and supported on $C_5 = C_4 \cup C_1$.
We will further have to specify a K\"ahler class and see if $L_5$ is stable.

We can also view this as a heterotic model. Namely we start with a reducible rank six bundle of the form
\be
V_6 \ = \ V_0 \oplus V_1 \oplus V_4
\ee
where the $V_i$ are the Fourier-Mukai transforms of the sheaves $L_Y^{-5/6}$, $Q$ and $L_4$ above.
Turning on $\bar N$ means that we replace
$V_1 \oplus V_4$ by a non-trivial extension $V_5$, whose Fourier-Mukai transform is $L_5$.
So the final configuration will be $V_0 \oplus V_5$. The original $V_1 \oplus V_4$ will be unstable,
but both $V_0$ and $V_5$ need to be slope stable of slope zero,
so that the sum $V_0 \oplus V_5$ is poly-stable.

The details for such constructions are explained in
\cite{Donagi:2008kj,Donagi:2009ra}. The homology class of $C_4$ is determined by
choosing a class $\eta \in H^2(S)$, and the class of the matter curve in $H^2(S)$ is then given by
\be
[\Sigma_{\bf 16}] = \eta - 4\, c_1(S).
\ee
We can further specify $L_4$ by finding a suitable algebraic
representative for its first Chern class. In order to get an ample
supply of such classes, it may be necessary to tune the complex
structure moduli to get Noether-Lefschetz classes. There is
enormous flexibility in this part of the construction so we will
not detail it any further.

It is useful to note some simple consistency constraints. We assume
that $L_Y^{5/6}$ does not contribute to the net chirality, as
happens under the usual condition that $c_1(L_Y^{5/6})$ is
topologically trivializable in the bulk of the compactification.
Suppose that we have $k_1$ generations of $\bt$ living on
$\Sigma_{\bf 16}$ and $k_2$ generations of $\bt$ from the bulk,
i.e. we have
\ba
\bt_{1/5}: & &  -\chi(V_4)\ =\ k_1 \eol[2mm]
\bt_{-4/5}: & &  -\chi(V_1)\ =\ k_2
\ea
Then we can deduce that
\be
\begin{array}{rcr}
\bfb_{2/5}: & &  -\chi(\Lambda^2 V_4) \ =\  k_2 - k_3\eol[2mm]
\bfb_{-3/5}: & &  -\chi(V_4 \otimes V_1) \ =\ k_1 + k_3 \eol[2mm]
{\bf 1}_{1}: & & -\chi(V_4 \otimes V_1^*)\ =\ k_1 - k_3
\end{array}
\ee
where $k_3 = \int_{\Sigma_{\bf 16}} c_1(Q)$.
We can adjust
$c_1(Q)$ to get various numbers of $SU(5)_{GUT}$ singlets charged
under $U(1)_{B-L}$. Now we will assume there are three generations
of $\bt_{1/5}$ on $\Sigma_{\bf 16}$. (Variations on this are
possible). If we further want three right-handed neutrinos $N$, and
one $\bar N$ (which will get eaten by the $U(1)_{B-L}$), then it appears we
should take $c_1(Q) \cdot [\Sigma_{\bf 16}] = 1$, although having
additional right-handed neutrinos may be useful.

Getting both $N$
and $\bar N$ on the same curve requires that the genus $g$ of the curve is at least
equal to one, and that the net number of generations
on the curve is between $\pm(g-1)$. It is easy to show using Riemann-Roch that one
cannot get chiral/anti-chiral pairs outside this range. Fortunately,
it is very easy to make the genus of
$\Sigma_{\bf 16}$ large by making $\eta$ moderately large.

Even inside this range, getting chiral/ant-chiral pairs is not
generic and requires tuning complex
structure moduli, i.e. it requires tuning a modulus $U$ appearing in
a Yukawa coupling $U N \bar N$. However $\bar N$ gets a VEV in the
final configuration we are interested in, so moduli such as $U$ will
be massive, and this is not unnatural. With one $N/\bar N$-pair, one modulus
$U$ pairs up with an $N$ when $\bar N$ gets a VEV,
so we should actually take $c_1(Q) \cdot [\Sigma_{\bf 16}] = 0$ if we want exactly
three right-handed neutrinos.

A reducible configuration like above is typically unstable. Recall
from the discussion above that we would like an expectation value
for $\bar N$, to lift dangerous $\btb$ bulk modes. This corresponds
to a gluing morphism pointing from $C_4$ to $C_1$. We would like to
argue that the configuration obtained by turning on such a VEV can
be made stable. The new rank five Higgs bundle $E$, given by the
extension of $p_{4*}L_4$ by $Q$, has
\be c_1(E)\ =\ c_1(p_{4*}L_4) + c_1(Q) \ = \  c_1(L_Y^{5/6}) \ee
In order to avoid lifting $U(1)_Y$ by closed string axions, the
cohomology class $c_1(L_Y^{5/6})$ should trivialize when we embed in
a compact model, and so $c_1(L_Y^{5/6})$ is orthogonal to the
K\"ahler class. This topological condition also
guarantees that there is no Fayet-Iliopoulos term for $U(1)_Y$. Then
the slope of $E$ is zero, and as discussed in section 3.2 of part I,
there are two natural Higgs sub-bundles, one of these being $Q$ (or
of course any line bundle that maps into $Q$). Proving stability
is a hard issue, and we will
not do so here. However for a configuration of the above type, there are only two natural
necessary conditions: the slope of $Q$ should be negative and
the slope of $Q(\Sigma_{\bf 16})$ should be positive. This can easily be arranged
for suitable choices of $Q$ and the K\"ahler class. We will choose
an explicit class below. The K\"ahler class would eventually be
determined by dynamical considerations, but that is beyond the scope
of our simple example, and we fix it by hand.

It remains to check for exotic bulk matter. We will calculate the spectrum
for $\vev{\bar N}=0$ and then check if the resulting fields may be lifted by a Yukawa coupling.
The net number of
$\bt_{-4/5}$'s in the bulk is given by $c_1(Q) \cdot c_1(K_S)$. The degree two homology lattice
of our del Pezzo surface is spanned by the hyperplane class $H$ and the exceptional $-1$-classes $E_i$ subject
to the relations:
\be
H\cdot H \ = \ 0, \qquad H \cdot E_i \ = \ 0, \qquad E_i \cdot E_j\ =\ -\delta_{ij}
\ee
We
take
\be
Q\ =\ \cO(E_1-E_2), \qquad \ L_Y^{5/6}\ =\ \cO(E_3 - E_4)
\ee
so there is no net matter in the bulk, and
$c_1(L_Y^{5/6})$ is trivializable. (Obviously variations on this theme are
possible, for instance we could have four net generations on
$\Sigma_{\bf 16}$ and one net anti-generation in the bulk, and then
lift the chiral/anti-chiral pairs using the gluing morphism). Note
that we can indeed make the slope of $Q$ negative, eg. by taking
\be
J \sim -c_1(K)
+\epsilon \, c_1(\cO(H-E_2)),
\ee
where $\epsilon$ is a real and
positive number. The the slope of $Q(\Sigma_{\bf 16})$ is given by
\be
\mu(Q(\Sigma_{\bf 16})) \ = \ -\,\epsilon + J \cdot [{\Sigma_{\bf 16}}]
\ee
It is easy to make this positive as well as satisfy $c_1(Q)\cdot [{\Sigma_{\bf 16}}]=0$
by making $\eta$
moderately large, eg. $\eta = 6c_1(S)$.
By exchanging $E_1 \leftrightarrow E_2$ we could
also reverse the signs, so that $N$ would get a VEV instead of $\bar
N$. At $\epsilon = 0$ the reducible configuration is poly-stable, so
$\epsilon = 0$ corresponds to a wall of marginal stability.

At any rate, with these choices we find the following cohomology groups for bulk matter
descending from the $\btb_{4/5}$:
\be
\begin{array}{lll}
  (\overline{\bf 3 \!}\,, {\bf 2})_{-1/6,4/5}: \quad  & H^0(Q^{-1} \otimes K)=0 & H^1(Q^{-1}) = 0 \\
  ({\bf 1}, {\bf 1})_{-1,4/5}: & H^0(Q^{-1} \otimes L_Y^{- 5/6} \otimes
K)=0 \quad & H^1(Q^{-1} \otimes L_Y^{- 5/6}) = 1 \\
  ({\bf 3}, {\bf 1})_{2/3,4/5}: & H^0(Q^{-1} \otimes L_Y^{ 5/6} \otimes
K)=0 & H^1(Q^{-1} \otimes L_Y^{ 5/6}) = 1
\end{array}
\ee
Furthermore, since $c_1(L_Y^{5/6})\cdot c_1(K_S) = 0$, we get the
same number of net generations for all the fields descending from
the $\btb_{4/5}$. We can also see this using that $Q \to Q^{-1}$ is
a diffeomorphism symmetry of the del Pezzo. Therefore, there is also
one $(\overline{\bf 3 \!}\,, {\bf 1})_{-2/3,-4/5} $ bulk mode and
one $({\bf 1}, {\bf 1})_{1,-4/5}$ bulk mode. These bulk modes cannot
get lifted because that would require a VEV for $N$ in the Yukawa
coupling $\bt_{-4/5} \btb_{-1/5} N_1$, and thus they will be part of
the Standard Model fields. Geometrically all these bulk modes
correspond to turning on off-diagonal components of the $SO(10)$
gauge field on $S$.

Now we need to consider the Yukawa coupling. Geometrically this
appears to correspond to the following: we take the generator $s \in
H^1(Q^{-1} \otimes L_Y^{\pm 5/6})$, restrict it to $\Sigma_{\bf
16}$, and then multiply it by the gluing morphism $f \in
\Hom_{\Sigma_{\bf 16}}(Q^{-1}, L_4^{-1} \otimes K_{C_4})$ corresponding to $\bar N$.
This composition
 yields a map
\ba
H^1(Q^{-1} \otimes L_Y^{\pm 5/6})\ \times\ \Hom_{\Sigma_{\bf 16}}(Q^{-1}, L_4^{-1} \otimes K_{C_4})
\ \to \ H^1(\Sigma_{\bf
16},L_4^{-1} \otimes L_Y^{\pm 5/6}\otimes K_{C_4}|_{\Sigma_{\bf
16}})\eol
\ea
If the image
$fs|_\Sigma$ is a non-zero, then the Yukawa coupling is non-vanishing and the dangerous
bulk mode is lifted. Computing this composition requires writing
down explicit generators and is therefore somewhat tedious, but it
ends up in the right place: we
know that $H^1_\Sigma= H^0(\Sigma_{\bf 16},L_4 \otimes L_Y^{\mp 5/6}
\otimes K_S|_{\Sigma_{\bf 16}})^\vee$ is non-vanishing and at least
three-dimensional, as there are three chiral generations localized
on $\Sigma_{\bf 16}$. Therefore generically we expect this lifting
to take place. The end result then is that the chiral and
anti-chiral fields get paired up, and we are left with a
three-generation model with one right-handed lepton (in the $ ({\bf
1}, {\bf 1})_{1,-4/5}$) and one right-handed up type quark (in the
$(\overline{\bf 3\!}\,, {\bf 1})_{-2/3,-4/5}$) living in the bulk,
and all the remaining Higgs and matter fields living on matter
curves. The fact that one $\bt$ multiplet in the bulk is incomplete
(with the remaining field in the multiplet living on a matter curve)
can be seen as a remnant of the no-go theorem of
\cite{Beasley:2008kw}.

\newsubsubsection{Emerging flavour structures}

In conclusion, just as in generic $SU(5)_{GUT}$ models, we do not
expect a general statement about problems with the spectrum, in
particular there is no issue with exotic states in the low energy
spectrum.

Then, what is the advantage of considering such degenerate models?
The point is that precisely due to the extra structure, the
situation is more interesting than in generic models, because such
models can possess various flavour structures. Indeed let us summarize
some of the structure we found in the above models.

We saw that for zero gluing VEV, the model
has both chiral and anti-chiral fields in the bulk which can not get
paired up, even though the mass term is allowed by all the gauge
symmetries. The fields in the bulk can only pair up with fields on
the matter curve after turning on the gluing VEV. The reason for
this seems to be that we need a coupling $\int_S {\rm Tr}(A \wedge A
\wedge \Phi)$ where $(A,A,\Phi)$ are all bulk fields, but no
suitable $\Phi$ is available. Perhaps this selection rule can be
rephrased in terms of a global symmetry. Furthermore we see that
these bulk fields appear in incomplete multiplets. This provides an
idea for simultaneously solving the $\mu$-problem and the doublet
triplet splitting problem without using extra $U(1)$ gauge
symmetries, by putting the Higgs fields $(H_u,H_d)$ in the bulk. We
will consider a model of this type in section \ref{FlavModel}.

The kinetic terms of matter localized in the bulk of a $7$-brane
scale differently with the volume from kinetic terms of matter
localized on a curve. Therefore in this model we get additional
flavour structure after canonically normalizing the kinetic terms,
see section \ref{DStructures}.

There are also
constraints on the holomorphic Yukawa couplings. We saw that a bulk
mode cannot appear in an up-type $\bt_{-4/5}\bt_{-4/5}\,\bfv$ Yukawa
coupling, as it has negative $U(1)_{B-L}$ charge and we cannot
compensate it by multiplying with a suitable number of $\vev{\bar
N_{-1}}$ VEVs. Thus our model gives rise to texture zeroes. For
more on the holomorphic couplings, see section \ref{FTextures}.

\newsubsection{$E_6$-models}

It is clear that our techniques can be used to construct a variety
of new models with degenerate covers. We briefly consider $E_6$
models, which leads one to study non-reduced covers. For simplicity
we will take the final GUT group to be $SU(5)_{GUT}$ and not break
it further to the Standard Model. The commutant of $E_6$ in $E_8$
is $SU(3)$. Thus from the spectral cover perspective, $E_6$ models
are a special case of a $3+2$ splitting of the $SL(5,{\bf C})$
cover, $C_5 = C_3 \cup C_2$, with $C_3$ a non-trivial covering and
$C_2 = 2S$.

The apparent $E_6$
gauge group is further broken to $SU(5)$ by turning on a rank one
sheaf $L_5$ on $C_5$, restricting to rank one sheaves $L_2$ on $C_2$
and $L_3$ on $C_3$, and a gluing morphism on the intersection. We
denote $C_3 \cap S$ by $\Sigma_{\bf 27}$, as it corresponds to the
curve where chiral fields in the ${\bf 27}$ of $E_6$ are localized.
Labelling the matter content by
representations of $SU(5)_{GUT} \times SU(2)
\times SU(3)\times U(1)_X$, we see that the matter fields are distributed in the
following way:
\be
\begin{array}{rcl}
  \Sigma_{\bf 27}: & &  (\bt,{\bf 3}, {\bf 1})_{2/5},\ (\bfb,\overline{\bf
3\!}\,,{\bf 1})_{4/5},\ (\bfb,{\bf 3},{\bf
2})_{-1/5} ,\ ({\bf 1},{\bf 3},{\bf 2})_{1}  \\[2mm]
  {\rm bulk}: & & (\bt, {\bf 1}, {\bf 2})_{-3/5} ,\ (\bfb,{\bf 1},{\bf
1})_{-6/5}
\end{array}
\ee
The main new feature compared to the previous subsection is that we
have a rank one sheaf $L_2$ supported on the non-reduced surface
$C_2 = 2S$. As discussed in section (2.5) of part I, 
$L_2$ could either be a rank two bundle on the
underlying reduced scheme $S$, or a sheaf which is non-trivial on
the first infinitesimal neighbourhood of $S$ (i.e. a nilpotent Higgs
VEV). Note further that $C_2 \cap C_3 = 2\, \Sigma_{\bf 27}$, where
$\Sigma_{\bf 27} = S \cap C_3$, so the gluing morphism can also be a
bit more intricate than in the previous example. Nevertheless the approach of building
the spectral sheaf from more elementary pieces works equally well here.

Models of this type have various interesting flavour structures. Apart from possible texture zeroes
associated to $U(1)_X$, we may also get texture zeroes associated to
$U(1)_{B-L}$ if $L_2$ is built as the extension of two line bundles.
There are also additional fields propagating in the bulk.  For a more detailed
discussion of flavour we refer to section \ref{Flavour}.

\newsubsection{IIa/M-theory models}

\subseclabel{IIaMModels}

We would briefly like to comment on the issue of bulk matter in
IIa/M-theory models. In this context, local models are described by
real Higgs bundles on a three-manifold $Q_3$ \cite{Pantev:2009de}.
We will only make some preliminary remarks about this interesting
subject, which really requires a more thorough investigation.

The data in these models consists of a complex flat connection ${\bf
A}$ on a real three-manifold $Q_3$, whose hermitian part satisfies a
harmonicity condition. The complex flat connection is allowed to
have singularities, which correspond to non-compact flavour branes.

The background flat connection breaks some initial gauge group $G'$
to a smaller gauge group $G$. In order to get chiral matter, we want
an unbroken gauge group $G$ on $Q_3$ and an infinitesimal
deformation $\delta {\bf A}$ which transforms in a chiral
representation of $G$. The wave functions of these modes depend on
the values of the background higgs field, and are peaked when the
higgs field vanishes. Thus to get a bulk mode, we want an enlarged
gauge group $G'$ to be `visible' on most of $Q_3$.

In the usual intersecting brane configurations, we have precisely
the opposite situation. The wave functions are confined to a small
region around the zeroes of the background higgs field, where $G'$
is unbroken, by a steep potential.

One may consider breaking $G'$ to $G$ on $Q_3$ by a real flat
connection. However in the conventional situation (no gluing VEVs),
this flat connection must be defined on all of $Q_3$. This can be
allowed when $Q_3$ is negatively curved, but we are usually
interested in models where $Q_3$ is positively curved (eg. $Q_3=S^3$
or $Q_3= S^3/\Gamma$ where $\Gamma$ is a finite group). In such
models $h^1(Q_3)=0$, and any background flat connection continuously
connected to the identity is actually gauge equivalent to zero. In
fact even when $Q_3$ is negatively curved, we still want
$h^1(Q_3)=0$ to avoid massless adjoint fields. Discrete Wilson lines
also do not help, as $\delta {\bf A}$ should also be globally
defined on $Q_3$ in this situation and therefore will also be gauge
equivalent to zero.

To avoid this, the basic idea is to consider spectral cover
components coinciding with the zero section outside a codimension
two subset $\Delta$, where one may have vertical components. Then
one may have $h^1(Q_3\backslash \Delta) \not = 0$ or
$h^1(Q_3\backslash \Delta, L) \not = 0$, where $L$ is a flat bundle
on $Q_3\backslash \Delta$. With a suitable parabolic structure along
$\Delta$, this could yield a valid gauge theory configuration, as
discussed in section 2.5 of part I and also in section 4 on the standard embedding.

The reason for considering such configurations stems from simple
phenomenological considerations. If chiral matter is localized at
points of $Q_3$, then matter interactions (in particular Yukawa
couplings) tend to be exponentially small. This can be seen in two
equivalent ways. We can in principle solve for the exact
wave-functions satisfying the linearized Hitchin's equations. They
are approximately though not exactly gaussian, and their classical
overlaps therefore tend to be small. Alternatively, since we are
computing $F$-terms, we can use the freedom to scale up the Higgs
field $\phi$, i.e. we consider the large angle approximation. In
this approximation the overlap integrals are zero, but there are
instanton corrections which generate the couplings, which again are
small. With bulk matter however, the wave functions overlap
classically and so Yukawa couplings would be present at tree
level and therefore of order one.

\newpage

\newsection{Flavour structures}
\seclabel{Flavour}

\newsubsection{Structures from D-terms}
\subseclabel{DStructures}

In generic $F$-theory models, the matter fields are localized on
matter curves. We have seen in this paper that we can get matter in
the bulk of the 7-brane by considering degenerate spectral covers.
No-go theorems can be circumvented by using all available
ingredients, in particular a non-zero gluing morphism. We have
argued that turning off the gluing morphism typically requires an
extra tuning of the K\"ahler moduli, and so is actually less
generic. Thus we may equally well incorporate bulk matter in
building realistic models. Later we will also see that certain
well-known heterotic models have $F$-theory duals with bulk matter.

Models with bulk matter automatically have additional structure in
the Yukawa couplings originating from the $D$-terms. The physical
Yukawa couplings $\hat Y$ can be expressed in terms of the
holomorphic Yukawa couplings $Y^0$ as
\be \hat Y_{\alpha\beta \gamma} \ = \ e^{{\cal
K}/2}{Y^0_{\alpha\beta\gamma}\over (K_\alpha K_\beta K_\gamma)^{1/2}
} \ee
where $K_\alpha$ denotes the K\"ahler metric for a chiral field.
Kinetic terms for modes localized in the 7-brane bulk and on a
7-brane intersection have different scaling behaviour with respect
to K\"ahler moduli. Furthermore, bulk modes descending from $A$ have
different scaling than bulk modes descending from $\Phi$. According
to \cite{Aparicio:2008wh} we have
\be K_A \sim {1\over t}, \qquad K_\Phi \sim 1, \qquad K_I \sim
{1\over t^{1/2}} \ee
where $t$ is the volume of the four-cycle, and ${\cal K}$ of course
also depends on $t$. The subscripts $(A,\Phi,I)$ denotes the origin
of the chiral field as a bulk mode from $A$ or $\Phi$, or from an
intersection. Thus we automatically get flavour structure. For
instance in the $SO(10)$ model discussed earlier, we had one
generation of $\bt$ in the bulk descending from $A$ and all other
matter and Higgs fields localized on curves. (This is not quite
right because the $\bt$ multiplet was incomplete, but let us ignore
that). Then we find that in the large volume limit, one generation
has Yukawa couplings which are parametrically larger than the other
two, and the bottom is suppressed compared to the top for the heavy
generation.

In fact we may also get models with bulk fields descending from
$\Phi$, as we saw for $E_6$ models. Thus we may contemplate the
following scenario: we take two generations of $\bt$ to be `mostly'
bulk modes, one from $A$ and one from $\Phi$. We say `mostly'
because as we have seen, when gluing VEVs are non-zero the
distinction between bulk modes and modes on matter curves becomes a
bit blurry. We take the remaining $\bt$ and all $\bfb_m$ matter to
live on matter curves. We will also assume a K\"ahler potential with
a large and a small size modulus. Then using equation (2.18) in
\cite{Aparicio:2008wh} one would get a flavour model of the form
\be M_u \sim \left(
               \begin{array}{ccc}
                 \epsilon^4 & \epsilon^3 & \epsilon^2 \\
                 \epsilon^3 & \epsilon^2 & \epsilon \\
                 \epsilon^2 & \epsilon & 1 \\
               \end{array}
             \right), \quad
M_d  \sim \left(
               \begin{array}{ccc}
                 \epsilon^3 & \epsilon^3 & \epsilon^3 \\
                 \epsilon^2 & \epsilon^2 & \epsilon^2 \\
                 \epsilon & \epsilon & \epsilon \\
               \end{array}
             \right)
\ee
up to an overall factor of $\epsilon^{1/2}$ which one could
hopefully offset against the numerical pre-factor. This is known to
be a decent approximation in the real world \cite{Haba:2000be}. However this does not
allow us to suppress proton decay, for which we will propose a
different mechanism.

The above scaling behaviours should probably not be taken too
literally. We have seen that we can degenerate a model with all
matter localized on curves to a model with some matter in the bulk.
Thus one should be able to smoothly interpolate between these
scaling behaviours. Furthermore the actual wave functions can get
quite complicated. The numerical approximation suggested in section
3.3 of part I may lead to a clearer picture.

\newsubsection{Textures from F-terms: holomorphic zeroes}

\subseclabel{FTextures}

Proton decay is an important issue in string GUT constructions, and
even in string MSSM constructions. The problem starts already with
the Yukawa couplings, as $R$-parity is not automatic.

One of the most plausible solutions to this problem, though not the
only one, is to realize $R$-parity as a remnant of an additional
light $U(1)_X$ symmetry. Such extra $U(1)$ symmetries forbid the
$\bt_m \cdot \bfb_m \cdot \bfb_m$ Yukawa coupling, as well as other
couplings. The most well-known of these is $U(1)_{B-L}$.

As is well-known, there is some tension in such a scenario. The
extra $U(1)$ gauge boson must acquire a mass above the weak scale to
avoid having been detected yet. But naively breaking the $U(1)$ should
reintroduce $R$-parity violating couplings, so the $U(1)$ cannot be
too heavy either. If the $U(1)$ gets a mass though the Higgs
mechanism, then it should be a few orders above the weak scale.
Whether this is possible depends sensitively on the values of the
soft terms, i.e. on the method of breaking and mediating
supersymmetry, and leads to a conflict with the Majorana neutrino
mass scenario. If instead the $U(1)$ boson gets a mass through a
St\"uckelberg coupling to K\"ahler moduli axions, then we still have
a good $U(1)$ symmetry in perturbation theory, violated only by
$M5$-instantons, and one may imagine the $U(1)$ boson to be heavier.
But this has its own problems, for instance the Majorana mass term
has to be induced by non-perturbative effects.

In this section, we consider the possibility that the extra $U(1)_X$
symmetry is broken through the VEV of a chiral field, rather than
through non-perturbative effects (which will still be present, but
small). The naive expectation that all $U(1)_X$ violating operators
get generated is in general not correct in supersymmetric theories.
Holomorphy may forbid certain couplings even when the $U(1)_X$ is
broken \cite{Nir:1993mx}. This mechanism leads to interesting
textures which are essentially remnant selection rules of the broken
$U(1)_X$ symmetry. By an artful assignment of the charges (but still
constrained by an embedding in $E_8$)\footnote{It is not
quite true that the $U(1)_X$ must be embedded in $E_8$, but here we
are only interested in the $U(1)_X$ charges of fields localized near the
GUT brane.}
one can forbid problematic couplings while allowing for others, thereby
addressing the tension one usually finds in such models.

Let us first consider engineering an extra $U(1)_X$ symmetry.
Unbroken symmetry generators correspond to cohomology classes
$\rho_\xi$ in $\mathbb{H}^0({\cal E}^{\, \bullet})$, i.e.
$\delb$-closed zero forms in $\Omega^0(Ad(E_8))$ that commute with
$\Phi$. Now for a generic Higgs bundle we would not get any such
endomorphisms. This follows from the usual arguments: a generic
spectral sheaf corresponds to a line bundle on a smooth and
irreducible surface, and the corresponding Higgs bundle is stable.
But stable Higgs bundles only have trivial endomorphisms, and so we
should look for a poly-stable Higgs bundle instead.

To get a non-trivial endomorphism, we therefore should make the
spectral cover reducible. However this condition is not sufficient
to enforce a massless $U(1)_X$, as it does not guarantee that the
Higgs bundle is reducible. For this, we need the rank of the map
$[\Phi,\cdot]$ to drop by one globally. As we have seen, the problem
is that $\Phi$ may have non-trivial Jordan structure at the
intersection of the reducible pieces. To get a massless $U(1)$, we
must also require that the Jordan structure or equivalently the
gluing morphism is trivial.

As explained in part I, turning off the
gluing morphisms corresponds to approaching a wall of marginal
stability. Let us consider the effective field theory at such a
wall, i.e. we consider the KK expansion around a reducible brane
configuration with zero gluing VEV. As discussed, the resulting
effective field theory is a version of the Fayet model. That is, in
the minimal case we have a $U(1)_X$ vector multiplet, a chiral field
$X$ charged under the $U(1)_X$, and a $D$-term potential
\be V_D \ =\ \half (\zeta_X + q_X |X|^2)^2 \ee
Turning on the Fayet-Iliopoulos parameter breaks the $U(1)_X$
symmetry.

Now let us consider the effect of $U(1)_X$ breaking on the Yukawa
couplings. When the $U(1)_X$ symmetries are explicitly broken by a
non-zero expectation value for $X$, at first sight one might think
there are no selection rules left on the couplings. However this is
not correct. It is clearest if we only have a single $U(1)_X$
charged field $X$, which we can take to have a positive sign for the
charge. Let us consider general higher order couplings in the
superpotential
\be W\ \sim\ W|_{X=0} + \Phi_\alpha \Phi_\beta \Phi_\gamma X +
\ldots \ee
After turning on a VEV for $X$, we get various couplings violating
the $U(1)_X$ symmetry. However for the superpotential to be
invariant, only holomorphic couplings that are negatively charged
by an integer multiple of $q_X$ can get generated in the effective
theory. (For dimension five and higher one should be careful that
the operator has a good $\vev{X}\to 0$ limit, but otherwise a
similar argument can be made). Such couplings may still be
generated non-perturbatively, but one may reasonably expect such
corrections to be small. This leads to definite textures in the
Yukawa couplings even after the $U(1)_X$ symmetry is broken, as
long as the effective Fayet-model is valid. This is the
`holomorphic zero' mechanism: holomorphy of the superpotential
prevents operators with the wrong sign of the charge from being
generated \cite{Nir:1993mx}. It has played an important role in
proving non-renormalization theorems in supersymmetric theories and
is clearly also present in string compactification, as has been
observed in the context of the heterotic string and $F$-theory in
for example
\cite{Ibanez:1994ig,Maekawa:2003qm,Tatar:2006if,Kuriyama:2008pv,Anderson:2010tc}.

Such textures tend to persist even if the low energy theory has
fields with both signs of the $U(1)_X$ charge, $X_+$ and $X_-$. This
is easily seen to be a consequence of the effective field theory, as
noted for example in \cite{Kuriyama:2008pv}. The symmetries allow
for a superpotential coupling
\be W \ \simeq \ (X_+ X_-)^2 \ee
and analogous higher order terms which have no reason to be absent.
They will generically get generated from integrating out the KK
modes. The $F$-term equations then forbid fields with both signs of
the $U(1)_X$ charges to get a VEV simultaneously. Consequently,
$U(1)_X$ textures with a definite sign will appear also in the more
general case.

Let us briefly discuss in which range of parameters we can trust
these textures. We have seen in section $3.2$ of part I that the
Fayet model itself can only be trusted for an infinitesimal
Fayet-Iliopoulos parameter.
This is expected from effective field theory, as the mass of the
$U(1)_X$ gauge boson should be parametrically small compared to the
KK scale in order to trust the Fayet model. Even this may not be
sufficient, it is the maximum that could be expected from the low
energy effective field theory at the wall. Depending on the
microscopic properties of the vacuum, new physics may come in below
that scale. In fact we already know that the breakdown of the Fayet
model happens strictly below the KK scale, because the VEV of $X$
depends on the Fayet-Iliopoulos parameter $\zeta_X$ which is a
function of the K\"ahler moduli, and the K\"ahler moduli are
dynamical modes stabilized below the KK scale.

Now let us simply fix the K\"ahler moduli by hand and ask what
happens when we increase the Fayet-Iliopoulos parameter. When the
mass of the $U(1)_X$ gauge boson, which is proportional to
$\vev{X}$, is not parametrically small compared to the KK scale, we
are required to reexpand around the true vacuum configuration. The
obvious configuration would correspond to deforming the brane
configuration by a finite deformation whose tangent vector is the
internal zero mode corresponding to the $4d$ field $X$. In other
words, we expand around a reducible brane configuration with
non-zero gluing morphism. If this configuration is stable, then we
can take this to be our new vacuum. But as we discussed in section
3.2 of part I, it is not guaranteed that this new configuration is
stable, despite the fact that it seems to be suggested by the Fayet model,
and in the generic case it would not be. If that is the
case, then we have to further deform the reducible brane
configuration to reach a true vacuum. Since the true vacuum is then
generically obtained by turning on modes with both signs of the
$U(1)_X$ charges (smoothing modes), the superpotential in the true
vacuum would not satisfy $U(1)_X$ selection rules.

At any rate, with these caveats we see that reducible configurations
of branes or spectral covers can give rise to Yukawa textures. For a
generic smooth spectral cover, we do not expect such Yukawa
textures. By Fourier-Mukai transform, our picture for reducible
brane configurations maps precisely to the construction of bundles
by extension, well-known in the heterotic string. Suppose that we
have a bundle $F$ given by the extension
\be 0 \ \to \ E_1\ \to\ F\ \to\ E_2\ \to \ 0 \ee
Applying the Fourier-Mukai transform, we find
\be
 0 \ \to \ {\sf FM}^1(E_1)\ \to\ {\sf FM}^1(F)\ \to\ {\sf FM}^1(E_2) \ \to \ 0 \ee
In particular, the support of ${\sf FM}^1(F)$ is simply the union of
the supports of ${\sf FM}^1(E_1)$ and ${\sf FM}^1(E_2)$, i.e a
degenerate cover. The only difference between ${\sf FM}^1(F)$ and
${\sf FM}^1(E_1)\oplus {\sf FM}^1(E_2)$ is the gluing morphism.

 The holomorphic zeroes arising from abelian gauge
symmetries are certainly not the only way to get $F$-term textures.
For instance it could happen that we get holomorphic zeroes because
the matter curves don't intersect. Or as we see in the next
subsection, certain superpotential couplings could be zero because
they involve bulk modes. However using abelian gauge symmetries is
fairly natural in string compactification and we now have a much
better understanding of the geometries that yield such textures. In
the next section we will combine it with bulk modes to
simultaneously address several phenomenological issues.

\newsubsection{A model of flavour, with Higgs fields in the bulk}

\subseclabel{FlavModel}

Now we would like to show how one can put the results of the
previous sections together and simultaneously address the problems
of $R$-parity, proton decay, the $\mu$-problem, and a crude flavour
hierarchy in $F$-theory GUTs. To be fair, apart from the Higgses we
will only consider the net amount of chiral matter. Nevertheless we
do not expect issues with anti-generations, because the only
unbroken gauge symmetry is the Standard Model gauge group, and
everything that is not protected by index theorems should get lifted
in a sufficiently generic model.

\newsubsubsection{Basic picture of the configuration}

The idea is to consider a $3+2$ split $C_5 = C_3 \cup C_2$ of the
$SL(5,{\bf C})$ spectral cover, so that the effective theory below
the KK scale has a certain extra $U(1)_X$ symmetry (often called
$U(1)_{PQ}$ in the $F$-theory literature
\cite{Heckman:2008qt,Marsano:2009wr}; see also
\cite{Hayashi:2010zp,Grimm:2010ez,Braun:2011zm} for global
aspects). As originally emphasized in \cite{Tatar:2006dc} in the
$F$-theory context, this symmetry can be used to forbid dimension
four and five proton decay and a $\mu$-term. Here we would like to
argue that we largely preserve $R$-parity and suppressed proton
decay while evading some of the problems when we break the $U(1)_X$
with a gluing VEV. The model is similar in spirit to
\cite{Tatar:2006if}. However, other than in previous work we will
not use this $U(1)_X$ symmetry to solve the $\mu$-problem. In
particular, this allows us to evade the problems with exotics
discussed in \cite{Marsano:2010sq,Dolan:2011iu}. Instead we will
use the observation in section \ref{SO(10)Models} about the
possibility of solving the $\mu$-problem and doublet/triplet
splitting problem by putting the Higgses in the bulk. This will
account for the Higgses being `different' from ordinary matter in
our scenario.

We will use an $SL(3)\times SL(2)$ spectral cover to break the
$E_8$ to $SU(6)$, but it will be more convenient to label the
fields by an $SU(5)_{GUT}\times SU(3)\times SU(2)\times U(1)_X$
subgroup of $E_8$. The matter fields in the $\bt$ and $\bfb$ split
up in the following way:
\be
\begin{array}{rcl}
  \bt: &  & (\bt, {\bf 1}, {\bf 2})_{-3/5} + (\bt,{\bf 3}, {\bf 1})_{2/5}
  \\[2mm]
  \bfb: &  & (\bfb,{\bf 1},{\bf 1})_{-6/5} + (\bfb,{\bf 3},{\bf
2})_{-1/5} + (\bfb, \overline{\bf 3\!}\,,{\bf 1})_{4/5}
\end{array}
\
\ee
We also have the $U(1)$-charged singlets:
\be ({\bf 1},{\bf 3},{\bf 2})_{1} + ({\bf 1},\overline{\bf 3\!}\, ,
{\bf 2})_{-1} \ee
We denote these singlets by $X^+$ and $X^-$.

From the representations, it is easy to read off the distribution of
matter in these models:
\be
\begin{array}{rcl}
  \Sigma_2= C_2 \cap S: &  & (\bt, {\bf 1}, {\bf 2})_{-3/5}  \\[2mm]
\Sigma_3 =C_3 \cap S: &  &  (\bt,{\bf 3}, {\bf 1})_{2/5} , \quad
(\bfb, \overline{\bf 3\!}\,,{\bf 1})_{4/5}
\\[2mm]
\Sigma_{32} = C_3 \cap C_2:  &   &  (\bfb,{\bf 3},{\bf
2})_{-1/5} , \quad ({\bf 1},{\bf 3},{\bf 2})_{1}   \\[2mm]
  {\rm bulk}: &  & (\bfb,{\bf 1},{\bf
1})_{-6/5}
\end{array}
\ee
The bulk modes can be understood as part of the adjoint of an
$SU(6)_{GUT}$ gauge group containing $SU(5)_{GUT}\times U(1)_X$ that
is left unbroken by our spectral cover. The fields on $C_3\cap C_2$
live in the fundamental ${\bf 6}$ of this $SU(6)_{GUT}$, the fields
on $C_3 \cap S$ live in the $\Lambda^2 {\bf 6} = {\bf 15}$, and the
fields on $C_2 \cap S$ live in the $\Lambda^3 {\bf 6} = {\bf 20}_v$
of $SU(6)_{GUT}$. The corresponding ALE fibration will have $SU(6)$
ALE singularities along $S$ because the resultant $R = b_0b_5^2 -
b_2 b_3 b_5 + b_3^2 b_4$ of a generic $SU(5)_{GUT}$-model vanishes
identically, so we may call this an $SU(6)$ model.

Let us make some more remarks on the matter curves. One may ask why
the $(\bfb,{\bf 3},{\bf 2})_{-1/5}$ is not localized on the zero
section $S$, since it is charged under the $SU(5)\times U(1)_X
\subset SU(6)$ gauge fields localized on $S$. In fact we can also
localize the computation of the spectrum of these fields on $S$.
Suppose that the sheets of $C_3$ are labelled by $\{
\lambda_1,\lambda_2,\lambda_3 \}$ with $\lambda_1 + \lambda_2 +
\lambda_3 = 0$, and the sheets of $C_2$ are labelled by
$\{\lambda_4,\lambda_5 \}$ with $\lambda_4 + \lambda_5=0$. We can
define a six-fold spectral cover $C_6$ whose sheets are labelled by
$\lambda_i + \lambda_j$, where $i=1,2,3$ and $j = 1,2$. Then the
computation of the $(\bfb,{\bf 3},{\bf 2})_{-1/5}$ can be localized
on $C_6 \cap S = \pi_* \Sigma_{32}$. Explicitly, if $C_3$ is given
by $c_0 \lambda^3 + c_2 \lambda + c_3 = 0$ and $C_2$ is given by
$a_0 \lambda^2 + a_2 = 0$, then $\pi_*\Sigma_{32}$ is given by
\be 0 \ = \ a_2^3 c_0^2 - 2 a_0 a_2^2 c_0 c_2 + a_0^2 a_2 c_2^2 +
a_0^3 c_3^2\ee
on $S$. This curve is singular however, and its normalization is
$\Sigma_{32}$. Analogous remarks apply to the spectrum of $({\bf
1},{\bf 3},{\bf 2})_{1}$, which is charged under the $U(1)_X$ gauge
field localized on $S$.

Now we will consider a scenario where the Higgses live in the bulk,
i.e. $H_u$ and $H_d$ should descend from $(\bfv,{\bf 1},{\bf
1})_{6/5}$ and its conjugate. In order to get all
$\bt_m\cdot\bt_m\cdot\bfv_h$ couplings of order one, we see that all
the $\bt$ matter comes from the $(\bt, {\bf 1}, {\bf 2})_{-3/5}$. We
further take $\bfb_m$ to descend from $ (\bfb, \overline{\bf
3\!}\,,{\bf 1})_{4/5} $. Clearly there are a number of variations on
our scenario, and we invite the reader to make his or her own.

It is easy to see that with this matter content, the $U(1)_X$ is
anomalous. There is an additional contribution from the net number
of $X^\pm$ fields, but this is not enough to cancel both the $\sum
q$ and $\sum q^3$ anomalies. This is not immediately deadly, because
$F$-theory permits axions and a Green-Schwarz mechanism, although it
may still imply constraints on the spectrum. (In $M$-theory models
by contrast, such anomalies would apparently doom the model
\cite{Pantev:2009de}). Furthermore we don't need this configuration
to be stable because we still want to turn on a VEV for $X^+$ and
break the $U(1)_X$. We will argue below that this matter content is
consistent with the index theorems, and so such models should exist.
In particular this addresses the question of stringy anomaly
cancellation, because this only depends on the net number of chiral
fields with given charges. It would be more satisfactory to check
that the actual matter content on the matter curves can be attained
without anti-generations, but such calculations are more involved.
As we explained in \cite{Donagi:2009ra} however, in local $F$-theory
models there is a landscape of solutions for the choice of flux on
the flavour branes (consisting of Noether-Lefschetz fluxes), and no
index theorem which protects pairs from lifting. So without further
calculations, naturalness demands we should assume there are no
anti-generations.

We will break the $SU(6)$ symmetry by a flux for $U(1)_X$ and a
gluing VEV. The $SU(5)_{GUT}$ symmetry will be further broken by
hypercharge flux. As the Higgses live in the bulk and the matter
fields live on curves, and we do not want to split the matter
multiplets, we will take the hypercharge flux through the matter
curves to be zero identically.

\newsubsubsection{Consistency}

Let us investigate if the spectrum can be arranged in this way. The
reader who is not interested in these technicalities may skip to
equation (\ref{AbsentCouplings}). It seems that $H_u$ and $H_d$ must
generically pair up, because the mass term $H_u H_d$ is a gauge
singlet. This is why frequently configurations are considered where
$H_u$ and $H_d$ are charged under the $U(1)_X$ symmetry so as to
forbid the $\mu$-term. But actually this expectation is not quite
true for bulk fields, as we saw for the $SO(10)$ models earlier. At
least for vanishing gluing VEVs, the index theorems for bulk modes
are stronger in the sense that they know about more than just the
net amount of chiral matter. As remarked earlier, this presumably
indicates the presence of a global symmetry under which bulk fields
are charged. The fields in the bulk transform as
\be ({\bf 2},{\bf 1})_{-1/2,-6/5} + ({\bf 1}, \overline{\bf
3\!}\,)_{1/3,-6/5} \ee
Thus we need to consider the rank two bundle on $S$ given by
\be (L_Y^{-1/2} \otimes L_X^{-6/5})   \oplus (L_Y^{1/3} \otimes
L_X^{-6/5}) \ \equiv\ (Q \otimes L_Y^{-5/6}) \oplus Q  \ee
We take $S$ to be a del Pezzo surface, and denote the hyperplane
class lifted from ${\bf P}^2$ by $H$ and the exceptional $-1$-curves
by $E_i$. Taking for example
\be
c_1(L_Y^{5/6})\  = \  \cO(E_3-E_4), \qquad
c_1(Q) \ = \  \cO(E_1-E_2),
\ee
both may restrict trivially to the
matter curves (which we still have to choose, we will do so below).
Then we have $H^1(S,Q^{\pm 1})=0$ and $ H^1(S,Q^{\pm 1}\otimes
L_Y^{-5/6}) =1$ as required. It follows that we simultaneously solve
the $\mu$-problem and the doublet/triplet splitting problem this
way.

We would now like to investigate the remaining fluxes and check that
everything is consistent. We can do this by mapping to an
elliptically fibered heterotic model $\pi:Z \to B_2$ with $B_2 = S$,
as the computations are slightly more straightforward that way. Here
we have rank three bundle $V$ and a rank two bundle $W$, both
constructed from spectral covers $(C_V,L_V)$ and $(C_W,L_W)$, such
that
\be c_1(V) \ =\ \pi^*c_1(Q^{-1} \otimes L_Y^{5/6}), \qquad \ c_1(W)
\ = \ \pi^*c_1(Q) \ee
That is, from the heterotic point of view we start with a rank six bundle of the form
$V_3 \oplus W_2 \oplus U_1$ where the subscripts indicate the rank. Eventually,
$V \oplus W$ will be replaced by a non-trivial rank five extension.
Then we have the following table:
\be\label{SU(6)HetMatterTable}
\begin{array}{rcl}
   \  (\bt,{\bf 3}, {\bf 1})_{2/5} &  &
   \chi(V)=0\\[2mm]
     \ (\bt, {\bf 1}, {\bf 2})_{-3/5}
  & & \chi(W) =-3 \\[2mm]
  \ (\bfb, \overline{\bf 3\!}\,,{\bf 1})_{4/5}
  & &  \chi(\Lambda^2V) = -3 \\[2mm]
   \   (\bfb,{\bf 3},{\bf
2})_{-1/5}&  &  \chi(V\otimes W)=0
\\[2mm]
 (\bfb,{\bf 1},{\bf 1})_{-6/5} & &
 \chi(\Lambda^2 W) = \chi(Q) =0\\[2mm]
  ({\bf 1},{\bf 3},{\bf 2})_{1}  & & \chi(V \otimes W^*) = N_X
\end{array}
\ee
We don't need to specify the net number of singlets, as tuning to
get vector-like $X^\pm$ pairs is fine for our purposes. Such tuning
moduli will be lifted after turning on the $X^+$ VEV. We do need
that $(N_{X^-}-N_{X^+})\geq -1$, because otherwise there will be
massless $X^+$ fields remaining after turning on an $X^+$ VEV. We
also need the genus of $C_2\cap C_3$ to be larger than the net number of
$X^\pm$ generations, because otherwise the
spectrum of $X^\pm$ is purely chiral by Riemann-Roch and we cannot tune to get
vector-like pairs. We will find below that actually
$(N_{X^-}-N_{X^+})=12$ for consistency. So the genus of $C_2\cap
C_3$ must be such that $g-1 \geq 12$. The matter curves are given
by
\be [\Sigma_3] = \eta_V - 3c_1 \qquad {\rm  and} \qquad [\Sigma_2] =
\eta_W -2 c_1 \in H^2(S). \ee
where $c_2(V) = \pi^*\eta_V$, $c_2(W) = \pi^*\eta_W$ and $c_1 =
c_1(T_S)$. The genus of $C_2 \cap C_3$ grows quickly if we take
$\eta_V$ and $\eta_W$ to be even moderately large.

Now let us consider the more refined information. Recall that in
order to get zero net generations in the bulk, we had $c_1(Q) \cdot
c_1 = 0$. We have \cite{Hayashi:2008ba}
\be \chi(\Lambda^2V) = (N-4)\chi(V) - \pi^*(\eta_V-3c_1)\cdot
\sigma_{B_2}\cdot c_1(V) \ee
where $\sigma_{B_2}$ is the section of the elliptic fibration.
Therefore lines one and three of (\ref{SU(6)HetMatterTable}) are
mutually consistent if
\be c_1(Q^{-1})\cdot \eta_V\ = \ c_1(Q^{-1})\cdot \Sigma_3\ =\ 3.
\ee
Here we simplified using the fact that $c_1(L_Y^{5/6})$ is
orthogonal to $c_1$ and to the matter curves.  Furthermore,
$[\Sigma_3] = \eta_V - 3c_1$ must be effective. We can satisfy this
for example by picking $\eta_V \sim n_Vc_1 +3 E_1$ with $n_V \geq 4$. We also find
\be
\begin{array}{ccc}
 \chi(V\otimes W) & = & {\rm rk}(W)\chi(V) + {\rm rk}(V)\chi(W) - c_1(V) \cdot \sigma \cdot
\pi^*\eta_W - c_1(W)\cdot \sigma \cdot \pi^*\eta_V \\
\end{array}
\ee
and therefore lines one, two and four of (\ref{SU(6)HetMatterTable})
are mutually consistent if
\be c_1(Q^{-1})\cdot \eta_W + c_1(Q)\cdot \eta_V \ = \ -9 \ee
This determines the flux through the remaining matter curves, and
the net number of $X^\pm$ fields. We find that
\be c_1(Q^{-1}) \cdot \eta_W\ =\ c_1(Q^{-1}) \cdot C_2\ =\ -6 \ee
and again $[\Sigma_2] = \eta_W - 2c_1$ must be effective. This can
also be achieved, eg. by picking $\eta_W \sim n_Wc_1 + 4 E_1 - 2 E_2$
with $n_W \geq 4$.
More minimal solutions can probably be obtained by picking a
slightly more complicated form for $c_1(Q^{-1})$. By comparing the
first two lines with the last line of (\ref{SU(6)HetMatterTable}),
and using the intersection numbers deduced above, we then predict
there are twelve net $X^-$ generations, four per generation of
matter. And we can check that the second and fifth lines of
(\ref{SU(6)HetMatterTable}) are mutually consistent, which also
works. Thus, any remaining anomalies can be cancelled by a
Green-Schwarz mechanism.

It remains to specify spectral line bundles on $C_V$ and $C_W$ so
that $\chi(V) = 0$ and $\chi(W) = -3$. Such constructions are
explained in \cite{Donagi:2008kj,Donagi:2009ra} and the result is
that by using the landscape of Noether-Lefschetz fluxes (and moderately
large $\eta$), one can get pretty much anything one wants.

\newsubsubsection{Phenomenological properties}

After these technicalities, let us now see what kind of structure we
can find in the interactions in our scenario. The $U(1)_X$ charge
assignments we have chosen were as follows:
\be \bt_m : -3/5, \qquad \bfb_m:\ 4/5,  \qquad \bfv_h:\ 6/5, \qquad
\bfb_h: \ -6/5, \qquad X^+: \ 1 \ee
Then the $U(1)_X$ symmetry forbids the following couplings
\be\label{AbsentCouplings} {\rm absent}: \quad \bfb_m  \bfv_h,
\qquad \bt_m \bfb_m \bfb_h , \qquad \bt_m \bfb_m \bfb_m , \qquad
\bt_m \bt_m\bt_m\bfb \ee
As discussed above, although the $\bfv_h \cdot \bfb_h$ term is
neutral, the $\mu$-problem and the doublet/triplet splitting problem
have already been solved. We now turn on a large VEV for the gluing
morphism $X^+$. This requires us to arrange the correct sign for the
Fayet-Iliopoulos parameter $\zeta_X$, and we can do this exactly as
in section \ref{SO(10)Models} by taking the K\"ahler class to be of
the form
\be
J \sim -c_1(K) + \epsilon\,c_1(\cO(H-E_2)).
\ee
As in section \ref{SO(10)Models}, actually proving stability would
require a lengthy analysis which we will not attempt, but one can
check the two natural necessary conditions. Then we generate down
type Yukawa couplings which are hierarchically suppressed compared
to the up-type couplings:
\be\label{YukawaHierarchy} \bt_m\bt_m\bfv_h + {\vev{X^+}\over
M}\,\bt_m \bfb_m \bfb_h  \ee
where the size of $X^+$ is set by the Fayet-Iliopoulos parameter,
and therefore by $\epsilon$. So we have a flavour hierarchy,
admittedly very crude, but one can try to improve this eg. by
considering the K\"ahler potential. Note that whereas the down type
couplings are localized at $\Sigma_2 \cap \Sigma_3$, the up-type
couplings are not localized at points but along $\Sigma_2$. We can
not get the couplings $\bfb_m \bfv_h$ or $ \bt_m \bfb_m \bfb_m $,
because they are still forbidden by the remnant selection rules
(i.e. the holomorphic zero mechanism). Thus, $R$-parity is still
preserved at this level, even though the $U(1)_X$ has been broken.
There are bilinear $R$-parity violating terms in the K\"ahler
potential, and they may affect neutrino oscillations, but this is
known not to be a serious issue. The $\mu$-term is also not
generated by a VEV for $X^+$.

We do generically get dimension five proton decay:
\be {\vev{X^+}\over M^2}\, \bt_m \bt_m\bt_m\bfb_m  \ee
but it is suppressed by $\vev{X^+}$. This suppression is due to the
suppression of the bottom Yukawa coupling (\ref{YukawaHierarchy}),
so by itself this doesn't quite solve the dimension five proton
decay problem. One could consider addressing this problem by
raising the masses of the squarks as in split SUSY models, but we
will not go into that here. The corresponding operator with $\bfb_m
\to \bfb_h$ is suppressed by $\vev{X^+}^3$ and may effectively be
ignored. It may be helpful to recall a simple relation between the
$U(1)$-charges of certain couplings, which arises from
\be (\bt_m\cdot \bt_m \cdot \bfv_h)(\bt_m\cdot \bfb_m\cdot \bfb_h) \
\simeq \ (\bt_m\cdot \bt_m\cdot \bt_m\cdot \bfb_m)(\bfv_h\cdot
\bfb_h) \ee
at the level of $U(1)$-charges. Since in our scenario the dimension
five proton decay operator is charged under a $U(1)$-symmetry and
the mu-term is neutral, it follows that either the top Yukawa or the
bottom Yukawa coupling (or both) must also be charged and vanish
when $\vev{X^+}=0$. Of course this is what we found above.

Finally, there is the question of the neutrinos. One might think
that it is straightforward to interpret the many $X^-$ fields as the
right-handed neutrinos $N$ in our scenario. We even get Majorana
neutrino masses:
\be {\vev{X^+}^2\over M^2}\, N^2  \ee
However it is not that simple because with our assignments, Yukawa
couplings of the form $\bfb_m\bfv_h X^-$ are perturbatively
forbidden by the $U(1)_X$ symmetry even after turning on a VEV for
$X^+$. Alternatively, the Weinberg operator
\be LH_uLH_u \ee
could be generated by integrating out Kaluza-Klein modes, but again
this is forbidden by the $U(1)_X$ symmetry. The only way to generate
it in our scenario seems to be through $M5$-brane instantons, which
can generate such terms when the $U(1)_X$ $G$-flux through the
$M5$-worldvolume is non-vanishing \cite{Donagi:2010pd}. Analogous
neutrino scenarios were considered in
\cite{Ibanez:2006da,Blumenhagen:2006xt}. Our model is slightly
different and yields an interesting twist on this, because Majorana
masses are actually allowed. Generically these non-perturbative
corrections give two contributions to the effective Weinberg
operator, and one should take the dominant contribution. On the one
hand it may generate the missing leptonic Yukawa couplings, which
then give rise to the Weinberg operator after integrating out $N$.
Since the Yukawa couplings are small, the Majorana mass term could
be much lower than usual. On the other hand $M5$-instantons may
generate the Weinberg operator directly as well. At least one can
say that in both cases, the resulting neutrino masses are guaranteed
to be small. This was much less clear in the scenario of
\cite{Ibanez:2006da,Blumenhagen:2006xt}, where the Majorana mass
term itself was generated by instantons.

We could consider several variations, such as allowing some of the
bilinear $LH_u$ terms, or allowing the $\bt_m\bfb_m\bfb_m$ coupling
for the third generation, to get more spectacular signatures. We can
also try to vary the complex structure so that one or two
generations of $\bt$ approximate a bulk mode. Then as pointed out
earlier, once the kinetic terms are properly normalized, one may get
further flavour hierarchies due to suppression by a K\"ahler
modulus.

Although the most dangerous sources of $R$-parity violation are
avoided, $R$-parity is not conserved. This is the basic prediction
of models of the above type (or indeed many models where $R$-parity
is part of a $U(1)$ gauge symmetry, since that gauge symmetry has to
be broken and it is often hard to do so without also breaking
$R$-parity). In principle one could preserve $R$-parity if we could
find a field $X$ with $R$-charge two to give a VEV to, but somehow
in models of the above type we only seem to find fields with charge
one.

$R$-parity violation has many interesting phenomenological
signatures. A summary of some of these, with emphasis on models
similar to the above where the most dangerous $\bt_m\bfb_m\bfb_m$
terms are still forbidden, can be found in \cite{Kuriyama:2008pv}.
The simplest prediction of course is that the LSP is not stable and
will decay. From a bottom-up perspective there is a bewildering
number of $R$-parity violating scenarios one could consider, but as
emphasized in \cite{Kuriyama:2008pv}, string compactification gives
a theoretical framework for a class $R$-parity violating models
using the holomorphic zero mechanism and embedding in $E_8$. Given
the difficulty of engineering an exact $R$-parity, and the
ubiquitousness of holomorphic zeroes due to abelian gauge symmetries
in string compactifications, we should perhaps take this seriously
as a possible testible prediction of string GUTs.

\newpage

\newsection{The standard embedding}

\seclabel{K3}

\newsubsection{General comments}

Some of the oldest and simplest examples of heterotic
compactifications are obtained by `embedding the spin connection in
the gauge connection,' i.e. by taking the non-trivial part of the
bundle $V$ for the left-movers to be the same as the tangent bundle
$TZ$ of the Calabi-Yau.

Now it is not hard to see that the standard embedding on an elliptic Calabi-Yau
gives rise to degenerate spectral
covers, with the structures that we have discussed extensively in part I and part II.
On non-singular elliptic fibers $E$, we have the short exact sequence
\be 0 \ \to\ T_E\ \to\ TZ|_E\ \to\ N_E\
\to\ 0 \ee
For any fibration, we have that $N_E$ is a trivial bundle, so when $Z$ is a three-fold we
have $N_E \simeq \cO_E \oplus \cO_E$. Furthermore away from the singular fibers, we have $T_E = \cO_E$.
Thus the spectral cover is supported on three copies of the zero section, plus possibly
some vertical components at the singular fibers.
Furthermore, the extension is generically non-trivial, so
there is also a non-reduced structure or equivalently a nilpotent Higgs VEV along the zero section.
And we expect (and will later show) that one gets non-trivial gluings along the intersection of the
horizontal and vertical components.

Although $T^3$-fibrations are much trickier, the analogous argument yields a similar degenerate spectral cover
picture for the dual $M$-theory models of standard embeddings. As we discussed in section \ref{IIaMModels},
this type of structure is even more interesting for IIa/$M$-theory models than for IIb/$F$-theory models.

Thus the Fourier-Mukai transform of the tangent bundle automatically leads one to consider gluings, bulk matter and so forth.
The fact that these issues already appear in the most prototypical of string compactifications seems to
us pretty remarkable. It is all the more curious that such structures have been largely ignored in the study of
models with branes.

In this section we consider the spectral
data for the tangent bundle of an elliptically fibered $K3$ surface in more detail.
This bundle has previously been considered in
\cite{Friedman:1997ih,Bershadsky:1997zv,Aspinwall:1998he,Lazaroiu:1997},
and the spectral cover was known to be of a very degenerate form. We
will summarize some of the findings of these papers, calculate the
Fourier-Mukai transform of the tangent bundle, and then
illustrate how to calculate the spectrum directly from the degenerate spectral cover.

\newsubsection{The tangent bundle of a $K3$ surface}

Let us first discuss the spectral data for the tangent bundle of a
$K3$ surface. Although one could use the general method for deriving
the spectral data of a linear sigma model, in the present case there
is a direct method which yields more insight. In general, for an
$SU(2)$ bundle with $c_2 = k$ on $K3$, the homology class of the
spectral cover is given by
\be [C]\ =\ 2 [\sigma_B] + k [E] \ee
and the spectral line bundle has degree $2k-6$. To find the spectral
cover for the tangent bundle, we proceed as in
\cite{Aspinwall:1998he}. Restricting the tangent bundle to $E$, we
have the exact sequence
\be\label{K3TangentSequence} 0 \ \to\ T_E\ \to\ T_{K3}|_E\ \to\ N_E\
\to\ 0 \ee
Now away from the singular fibers, we clearly we have $T_E = N_E =
\cO_E$, so the spectral cover is given by twice the zero section,
plus perhaps some of the singular fibers. But we also know the
homology class of $C$. Given that $c_2 = 24$ we see that the full
spectral cover is given by twice the zero section together with the
24 vertical fibers. Since vertical components do not occur when $E$
is regular, we identify the vertical fibers of the spectral cover
with the twenty-four singular elliptic fibers.

We further want to know if the spectral sheaf on the zero section
corresponds to a rank two bundle or a non-trivial sheaf on the first
infinitesimal neighbourhood. This is the question whether the above
sequence (\ref{K3TangentSequence}) splits. The short exact sequence
(\ref{K3TangentSequence}) leads to the long exact sequence
\be
\begin{array}{ccccccccc}
  0 & \to & H^0(\cO_E) & \to & H^0(TK3|_E)& \to & H^0(\cO_E) & &
  \\[3mm]
   & \to  & H^1(\cO_E) & \to & \ldots &
   &  &  &
\end{array}
\ee
and the sequence splits when the coboundary map vanishes. Now the
coboundary map is also precisely the Kodaira-Spencer map. The
Kodaira-Spencer map is the derivative of the period map ${\bf P}^1_b
\to {\bf P}^1_\tau$, so the zeroes correspond to the branch points
of the period map. We can compute this number by a Riemann-Hurwitz calculation.

The period domain can be thought of as a Riemann sphere ${\bf P}^1_\tau$ with three-special points,
which we call $w=0,1$ and $\infty$.
At infinity the elliptic curve is nodal. At $w=0$ the elliptic curve has a
${\bf Z}_4$ symmetry, and there
is a ${\bf Z}_4$ monodromy around this point. Finally at $w=1$ the elliptic
curve has a ${\bf Z}_6$ symmetry, and there
is a ${\bf Z}_6$ monodromy around this point.

The period map ${\bf P}^1_B \to {\bf P}^1_\tau$ has degree $24$, as
there are $24$ singular fibers. Over $w=0$, the $24$ sheets meet in
$12$ pairs, and the ${\bf Z}_4/{\bf Z}_2 ={\bf Z}_2$ monodromy
interchanges the two sheets in each pair. (The ${\bf Z}_2$ subgroup
which corresponds to inversion on the elliptic curve acts trivially
on these sheets). Over $w=1$ the $24$ sheets meet in eight triples,
and the ${\bf Z}_6/{\bf Z}_2 ={\bf Z}_3$ monodromy acts on the
three sheets in each triple. The remaining ramification points are
the ones whose number we want to calculate (see \cite{CoxDonagi},
proposition 3.3).

Now the Riemann-Hurwitz formula says that
\be
\chi({\bf P}^1_B) \ = \ -2 \ = \ \chi({\bf P}^1_\tau) \times 24 + {\rm Ram}
\ee
and we have
\be
{\rm Ram} \ = \ 12\times (2-1) + 8 \times (3-1) + R \times (2-1)
\ee
where $R$ denotes the remaining ramification points. So calculation
shows that there are $R=18$ branch points, i.e. there are eighteen
points on the base where we get $\cO_E\oplus \cO_E$ instead of the
non-trivial extension, which we denote by $F_2$.

In heterotic models with $SU(2)$-holonomy bundles, there are two
interesting cohomology groups one could understand. Embedding the
spin connection in the gauge connection yields a six dimensional
theory with $E_7$ gauge group, $h^1(TK3)=20$ half-hypermultiplets
in the ${\bf 56}$ of $E_7$, and $2\times 45 = 90$ moduli. Here we
want to understand how the ${\bf 56}$ matter fields are realized
for the degenerate spectral cover dual to the tangent bundle, i.e.
in the $7$-brane picture.

The number of ${\bf 56}$ matter multiplets is counted by $H^1(TK3)$,
which just counts the number of complex structure deformations of
$K3$. Using the Parseval theorem for the Fourier-Mukai transform,
which says that
\be \Ext^p(V_1,V_2) \ = \ \Ext^p({\sf FM}^1(V_1),{\sf FM}^1(V_2))
\ee
we find that
\be\label{K3ExtGroup} \dim \Ext^1(i_{B*}\cO_{{\bf P}^1}, {\cal L})\
=\ 20 \ee
Here we have switched to the convention where the Poincar\'e sheaf
is normalized so that ${\sf FM}^1(\cO_{K3})= i_{B*}\cO_{{\bf
P}^1}$, in order to avoid factors of the canonical bundle $\cO_{{\bf
P}^1}(-2)$ from floating around. To understand these deformations,
clearly we will only need the behaviour of the spectral cover near
the zero section, so we may as well perform a normal cone
degeneration and study the resulting spectral cover in the total
space of $\cO(-2)_{{\bf P}^1}$ over the base ${\bf P}^1$. Let us
temporarily assume that the spectral cover on which ${\cal L}$ is
supported is smooth and generic. A generic curve in the linear
system $2S + \pi^*\eta$, with $\eta$ consisting of twenty-four
points on the base, is of the form
\be g_{24} s^2 + f_{20} = 0 \ee
and intersects the zero section precisely twenty times. The $\Ext$
group (\ref{K3ExtGroup}) counts the gluing morphisms that one can
turn on at each such intersection point, so from this point of view
the number of multiplets in the ${\bf 56}$ is twenty, precisely as expected.

Now we consider the degenerate cover corresponding to the tangent
bundle. As we discussed above, the Higgs field must be generically
nilpotent, but there are twenty-four vertical components and
eighteen points where the Higgs field vanishes. So after normal cone
degeneration, our Higgs field should be of the form
\be\label{TK3Phi} \Phi\ \sim\ \left(
  \begin{array}{cc}
    0 & f_{18}/g_{24} \\
    0 & 0 \\
  \end{array}
\right) \ee
acting on $E = \cO(2) \oplus \cO(-2)$.

\newsubsection{Derivation of the Fourier-Mukai transform}

We want to understand $\Ext^1(i_*\cO, {\cal L})$ (the number of
${\bf 56}$s) directly for the degenerate cover dual to the tangent
bundle. Of course we already know that we are going to get twenty
generators. But we would like to illustrate how to explicitly
calculate with degenerate $7$-brane configurations and identify the
actual deformations corresponding to the matter fields. To do the
calculation, we first need a more precise description of ${\cal
L}$. We will try to break ${\cal L}$ into several pieces.

Let us first try to find a global version of
(\ref{K3TangentSequence}) which also holds at the singular fibers.
Denote by $T_\pi$ the tangent fibers to the projection map
$\pi: K3 \to {\bf P}^1_B$. Then we claim that
\be\label{GlobalPiSeq}
0 \ \to \ T_\pi \ \to \ TK3 \ \mathop{\longrightarrow}^{d\pi}  \ \pi^*T_B \otimes I_{24}\ \to 0
\ee
Here $I_{24}$ is the ideal sheaf of the singular points on the $24$
nodal fibers. This sequence is clearly correct generically, so the
main thing to check is that it also works at the singular points on
the nodal fibers. We can do this by writing the map explicitly in
local coordinates near these points. It is given by an equation
\be xy \ = t
\ee
where $t$ is a local coordinate on the base. The projection map is
given by $\pi = xy$. Note that the total space is smooth and that
the singularity only appears by looking at the individual fiber
$\pi^{-1}(0)$.

Now we can locally identify the tangent bundle of the $K3$ with
\be
TK3 \ = \ {\bf C}[x,y]\vev{\del_x,\del_y}
\ee
and $d\pi$ is given by
\be
d\pi: (\del_x,\del_y) \ \to \ (y\del_t, x \del_t)
\ee
Note that this map is not surjective at $x=y=0$; this is why we
have the ideal sheaf $I_{24}$ appearing on the right of
(\ref{GlobalPiSeq}). For subsequent use, we also note that $T_\pi$
is generated by $x\del_x - y \del y$.

We further claim that we have an injective map $\pi^*T^*_B \to
T_\pi$ by contracting with the Poisson structure $\omega^{-1}$.
Again this is clear generically but we need to check what happens
at the singular point on the nodal fibers. Using the equation
above, $\pi^*T^*_B$ is generated by $dt$, and the holomorphic
$(2,0)$ form is given by $\omega = dx \wedge dy$. Thus we get the
map
\be
dt = xdy+ydx\ \ \mathop{\longrightarrow}^{\omega^{-1}}\ \ x\del_x - y \del_y
\ee
As noted above $T_\pi$ is generated by $x\del_x - y \del y$,
therefore contraction with $\omega^{-1}$ extends to an isomorphism
$\pi^*T^*_B \cong T_\pi$ on the whole $K3$, even on the nodal
fibers. Thus we have established a short exact sequence
\be\label{EllK3SES}
0 \ \to \ \pi^*T^*_B \ \to \ TK3 \ \mathop{\longrightarrow}^{d\pi}  \ \pi^*T_B \otimes I_{24}\ \to 0
\ee
where of course $T^*_B = \cO(-2)_{{\bf P}^1}$ and $T_B = \cO(2)_{{\bf P}^1}$

Now we can apply the Fourier-Mukai functor ${\sf FM}^\bullet$ to
this short exact sequence and deduce the structure of ${\cal L} =
{\sf FM}^1(TK3)$. We get the long exact sequence
\be\label{LExtension}
\begin{array}{ccccccccc}
  0 & \to & 0 & \to & 0 & \to & 0  & &
  \\[3mm]
   & \to  & i_{B*}T^*_B & \to & {\cal L} & \to & {\cal K}
    & \to & 0
\end{array}
\ee
Thus we see that ${\cal L}$ is given by an extension of ${\cal K}$
by $i_{B*}T^*_B$, where ${\cal K}= {\sf FM}^1(\pi^*T_B \otimes
I_{24})$ needs to be understood in more detail.

Let us denote the $24$ singular points on the nodal fibers by $p_i$. Then we
have another short exact sequence:
\be
0 \ \to \ \pi^*T_B \otimes I_{24} \ \to \ \pi^*T_B\ \to \ \pi^*T_B|_{24 p_i} \ \to \ 0
\ee
Applying the Fourier-Mukai transform, away from the singular points we get
\be\label{KExtension}
\begin{array}{ccccccccc}
  0 & \to & 0 & \to & 0 & \to & \oplus_{i=1}^{24} \cO_{E_i}  & &
  \\[3mm]
   & \to  & {\cal K} & \to & i_*T_B & \to & 0
    &  &
\end{array}
\ee
where $b_i =\pi(p_i)$. In other words, away from the singular
points, ${\cal K}$ is just the extension of $\cO(2)$ on the base by
$24$ vertical components. We have not determined what the sheaves
supported on the vertical components are exactly at the singular
points $p_i$. In fact this does not really matter for the
computation of $\Ext^1(i_*\cO, {\cal L})$, which is localized on
the zero section, away from the $p_i$. It does matter for our
calculation however if the extension above is non-trivial -- that
is, if the gluing morphism at the intersection of the horizontal
and vertical components is non-zero. From our discussion in part I,
we expect that the gluing morphism is non-zero, because turning off
the gluing morphism is singular and would lead to a new branch (the
small instanton transition). We will now check this explicitly.

The fiber of ${\cal L}$ at $b_i$ is given by
\be
{\sf FM}^1(TK3)|_{b_i} \ = \ H^1(E_i, TK3|_{E_i} \otimes \cO(b_i - \sigma)|_{\sigma =b_i} )
\ = \ H^1(E_i, TK3|_{E_i})
\ee
where $E_i$ is the fiber over $b_i$. The question is whether this
is rank one (so that the gluing morphism is non-zero, and we get a
line bundle) or whether this is rank two (so that the gluing
morphism is zero).

To get $H^1(E_i, TK3|_{E_i})$, we can try to restrict our short
exact sequence (\ref{EllK3SES}) to the singular fiber $E_i$ and
take cohomology. This doesn't immediately work because the
restriction is not exact. We get
\be\label{ResDPiSeq}
0 \ \to \ ker(d\pi|_{E_i}) \ \to \ TK3|_{E_i} \ \to \ I_{p_i} \ \to \ 0
\ee
but $ker(d\pi|_{E_i})$ is not quite the same as $(ker\,
d\pi)|_{E_i}= \pi^*T^*_B|_{E_i} = \cO_{E_i}$. The failure again
happens at the singular point, and we may use a local calculation
to deduce $ker(d\pi|_{E_i})$.

Using our previous model $xy=t$, we see that $ker(d\pi|_{E_i})$ is
generated by
\be
ker(d\pi|_{t=0}): \quad a\del_x + b \del_y\ | \ ay + bx = 0, \, xy=0
\ee
Away from $x=y=0$ this is one-dimensional. But at $x=y=0$ this is
two-dimensional. In other words, denoting the normalization map of
the nodal ${\bf P}^1$ by $\nu$, we have $ker(d\pi|_{E_i})=
\nu_*\cO_{{\bf P}^1}(k)$ for some $k$. In fact, since we have the
natural inclusion
\be
\cO_{E_i} = (ker \, d\pi)|_{E_i}\ \   \hookrightarrow\ \  ker(d\pi|_{E_i})
\ee
we see that actually $ker(d\pi|_{E_i}) = \nu_*\cO_{{\bf P}^1}(0)$.

So we can now take cohomology of (\ref{ResDPiSeq}). In the associated
long exact sequence
we will encounter the cohomologies
$H^n(\nu_*\cO_{{\bf P}^1}(0))$ and $H^n(I_{p_i})$. We can calculate
$H^n(\nu_*\cO_{{\bf P}^1}(0))$ from the Leray sequence
associated to the normalization map $\nu$. It is not hard to see that
$H^0(\nu_*\cO_{{\bf P}^1}(0))=1$ and $H^1(\nu_*\cO_{{\bf P}^1}(0))=0$.
Similarly, using the short exact sequence
\be
0 \ \ \to I_{p_i}\ \to\ \ \cO_{E_i} \ \to \ \cO_{p_i} \ \to \ 0
\ee
we calculate that $H^0(I_p) = 0$ and $H^1(I_p) = 1$. Then the long exact sequence
associated to (\ref{ResDPiSeq}) gives us
\be
\begin{array}{ccccccccc}
  0 & \to & 1 & \to & H^0(TK3|_{E_i}) & \to & 0  & &
  \\[3mm]
   & \to  & 0 & \to & H^1(TK3|_{E_i}) & \to & 1
    &  &
\end{array}
\ee
Hence we find that $H^1(TK3|_{E_i})=1$. So ${\cal L}|_{b_i}$ is a line bundle
and the gluing morphism is non-zero, as promised.

To summarize, we found that ${\cal L} = {\sf FM}^1(TK3)$ can be built explicitly
from two extension sequences, given in (\ref{LExtension}) and (\ref{KExtension}).
Furthermore, we checked that the
gluing morphisms appearing in (\ref{KExtension}) are non-zero. This matches well with
our earlier expectations.

\newsubsection{Calculation of the spectrum}

Given this explicit presentation,
we can now proceed to find $\Ext^1(i_*\cO,{\cal L})$. First we compute
$\Ext^1(i_*\cO,{\cal K})$ using (\ref{KExtension}). We get the long exact sequence
\be
\begin{array}{ccccccccc}
  0 & \to & 0 & \to & \Ext^0(i_*\cO,{\cal K}) & \to & \Ext^0(i_*\cO,i_*T_B)  & &
  \\[3mm]
   & \to  & \oplus_{i=1}^{24} \Ext^1(i_*\cO,\cO_{ E_i}) & \to & \Ext^1(i_*\cO,{\cal K}) & \to & \Ext^1(i_*\cO,i_*T_B)
    & \to & 0
\end{array}
\ee
where we used that $\Ext^p(i_*\cO,\cO_{E_i})=0$ for $p=0$ and
$p=2$. Now $\Ext^0(i_*\cO,i_*T_B) = H^0(\cO_{{\bf P}^1}(2))$ which
is three-dimensional, and $\oplus_{i=1}^{24} \Ext^1(i_*\cO,\cO_{
E_i})$ consists of $24$ gluing deformations. The coboundary map is
given by taking a generator of $H^0(\cO_{{\bf P}^1}(2))$,
restricting to $b_i$ and multiplying with the gluing VEV of ${\cal
K}$ at $b_i$. The gluing deformations in the image of this
coboundary map can be removed by symmetries. Thus only $24-3=21$ of
the gluing deformations in $\oplus_{i=1}^{24} \Ext^1(i_*\cO,\cO_{
E_i})$ are honest deformations that inject to $ \Ext^1(i_*\cO,{\cal
K})$, and we also get $\Ext^0(i_*\cO,{\cal K})=0$. We further have
\be
\Ext^1(i_*\cO,i_*T_B)\ \cong\ H^1(\cO_{{\bf P}^1}(2)) \oplus H^0(\cO_{{\bf P}^1}(0))
\ee
which yields zero non-abelian bundle deformations and one nilpotent
deformation on the base. Thus altogether we find that
$\Ext^1(i_*\cO,{\cal K})$ consists of $22$ first order
deformations.

Finally we apply $\Ext$ to the exact sequence in (\ref{LExtension}). We find
\be
\begin{array}{ccccccccc}
  0 & \to & 0 & \to & \Ext^0(i_*\cO,{\cal L}) & \to & 0  & &
  \\[3mm]
   & \to  & \Ext^1(i_*\cO,i_*T^*_B) & \to & \Ext^1(i_*\cO,{\cal L}) & \to & \Ext^1(i_*\cO,{\cal K})
    &  &  \\[3mm]
    & \to & \Ext^2(i_*\cO,i_*T^*_B) & \to &  \Ext^2(i_*\cO,{\cal L}) & \to & 0 & &
\end{array}
\ee
The result now depends on the rank of the coboundary map, which is
given by composing with the extension class in $\Ext^1({\cal K},
i_*T^*_B)$ that was used to build ${\cal L}$. This extension map is
non-zero except at eighteen points on the base, and these points
are separate from the intersections with the vertical fibers where
$\Ext^1(i_*\cO,{\cal K})$ are localized, so we expect that the rank
of the coboundary map is maximal. Now we have
\be
\Ext^2(i_*\cO,i_*T^*_B) \ \cong \ H^0(\cO_{{\bf P}^1}(2))^*
\ee
which is three dimensional, and similarly $\Ext^1(i_*\cO,i_*T^*_B)$ is one dimensional. This gives us that
$\Ext^1(i_*\cO,{\cal L})$ has dimension $22-3+1 =20$, as required.

Since we are considering compactification on $K3$, we have extra
supersymmetry, and our chiral fields in $\Ext^1(\cO_S,{\cal L})$ get
paired with chiral fields in $\Ext^1({\cal L},\cO_S)$ into
hypermultiplets. We have
\be \Ext^1({\cal L},\cO_S)\ \simeq\ H^1(T^*K3) \ee
so these can be interpreted as deformations of the cotangent bundle.
In the spectral cover picture these deformations are the duals of
the ones found before.

 The second cohomology group we could check is
$\Ext^1({\cal L},{\cal L})$, or equivalently the moduli of the
spectral sheaf. The moduli space is hyperk\"ahler and admits a
fibration by tori. The base is given by deformations of the spectral
cover, and the fiber corresponds to the Jacobian. The base and the
fiber have the same complex dimension. For a generic $SU(2)$ bundle
with $c_2 = k$, the genus of the spectral curve (and hence the
dimension of the Jacobian) is $2k-3$, which yields $45$ for $k=24$,
and the degree of the spectral sheaf is $g-3 = 42$. The Mukai moduli
space of sheaves on $K3$ has only one component, so we should get
the same dimension for our non-reduced spectral cover dual to the
tangent bundle. This yields the quaternionic dimension
\be \dim H^1({\rm End}(TK3)) \ = \ \dim \Ext^1({\cal L},{\cal L}) \ =
\ {45} \ee
Again we would like to understand these deformations explicitly for
the degenerate cover. We could calculate this explicitly as above,
but it is simpler to guess in this case. Using the embedding
$j:{\bf P}^1 \hookrightarrow C$
we get
\be
({\rm moduli\ that\ keep\ } {\bf P}^1\ {\rm fixed}) \ \to H^0(N_{C/K3}) \ \to \ H^0({\bf P}^1,\cO(20))
\ee
The deformations on the right are obtained from pulling the normal
bundle $N_{C/K3}$ back to ${\bf P}^1$. Since the class of $C$ has
intersection number $20$ with the ${\bf P}^1$, this gives us
$j^*N=\cO(20)$. These deformations correspond to smoothing
deformations, and there are 21 of them. The moduli on the left
correspond to moduli that leave the embedding ${\bf P}^1
\hookrightarrow C$ fixed. They correspond to moving the 24 fibers.
So in all we get $45$ moduli for deforming the spectral cover $C$,
as expected. Similarly we get one line bundle modulus from the line
bundle on each of the $24$ vertical fibers, and $24-3=21$
independent gluing VEVs at the intersections.

It is interesting to compare this to the sheaf ${\cal I}_{24}$
corresponding to 24 pointlike instantons \cite{Aspinwall:1998he}.
Clearly on the generic elliptic fiber away from the 24 point-like
instantons we have
\be {\cal  I}_{24}|_E \ =\ \cO_E \oplus \cO_E \ee
and since $c_2 = 24$, the spectral cover for ${\cal I}_{24}$ is also given
by twice the zero section and 24 elliptic fibers (though these need
not coincide with the singular fibers). The difference with the
tangent bundle is that this has an ordinary Higgs field VEV with
only trivial Jordan structure, and zero gluing VEV at the
intersection with the vertical fibers associates to the point-like
instantons. This difference manifests itself in how the moduli are
realized and in a jump in the number of matter fields. These computations
are much simpler than for the tangent bundle and left as an exercise
(partially done in \cite{Aspinwall:1998he}).

\newpage

\newsection{Linear sigma models}

\seclabel{LSM}

There are many realizations of Grand Unified Models in string
theory. Although it is not quite precise, one could say that
morally all these different realizations are dual to each other.
Here we would like to focus on the connections between three
realizations for which currently the most powerful techniques are
available: $F$-theory, large volume heterotic models, and heterotic
Landau-Ginzburg orbifolds. As we briefly reviewed, $F$-theory is
weakly coupled in the regime where the $8d$ heterotic string
coupling is large. Landau-Ginzburg orbifolds are special models
which are valid in a regime where the heterotic string coupling and
K\"ahler moduli are small. We can hope to learn about each of these
realizations by comparing with the others.

These questions have some relevance for phenomenology. We do not have a
non-perturbative formulation, and each weak coupling description is in
principle only valid for infinitesimal values of the coupling -- in $F$-theory,
these couplings are inverse volumes in Planck units. An old argument of Dine and
Seiberg \cite{Dine:1985he} essentially guarantees that we cannot
find a string vacuum with all moduli stabilized in perturbation
theory. An exception to this argument would be if we could define a
large $N$ expansion, which requires an infinite number of vacua --
this question has not yet been settled\footnote{An example of a runaway mode that
is absent for zero coupling was studied in \cite{Bena:2006rg}. Also
interesting in this regard is the warped deformed conifold. It
exhibits a `K\"ahler' mode that is parametrically light compared to
the KK scale \cite{Gubser:2004qj}, even though compactification on
the unwarped $T^*S^3$ does not yield any such light modes, i.e.
turning on the flux is not a small perturbation.}. On
the other hand, one
could not simply disregard the evidence from perturbation theory. A
better understanding of $F$-theory away from the large volume limit
could help illuminate possible qualitative issues with vacua
constructed using a perturbative expansion. Of course this goes both
ways, and one can also learn about strong coupling behaviour in the
heterotic string.

In order to set up such comparisons, we would like to be able to translate
between the linear sigma model description of bundles, and the spectral
cover description of bundles.
The main purpose of this section is to find an explicit algorithm
for producing the spectral sheaf associated to a monad on an
elliptically fibered Calabi-Yau $Z$. Before we specialize to monads
however, it will be useful to make some remarks that apply more
generally.

\newsubsection{General comments on bundles over elliptic fibrations}

\subseclabel{BundleGeneralities}

Given any bundle $\tilde V$ on our elliptic Calabi-Yau $Z$, there is
a natural `evaluation' map
\be\label{CanonicalBundleMap}  \Psi: \ \pi^*\pi_*\tilde V \ \to \
\tilde V \ee
where $\pi:Z \to B_2$ is the projection to the base. Let us describe
this map. Given a (sufficiently small) open set $U\subset Z$, the
local sections of $\pi^*\pi_*\tilde V$ are given by
\be \Gamma(U,\pi^*\pi_*\tilde V)\ =\ \Gamma(\pi^{-1}\pi(U),\tilde V)
\ee
Since $\pi^{-1}\pi(U)$ is obviously bigger than $U$, global sections
of $\tilde V$ over $\pi^{-1}\pi(U)$ are clearly a subset of global
sections of $\tilde V$ over $U$. We can always restrict a global
section over $\pi^{-1}\pi(U)$ to get a global section over $U$ (but
not vice versa). This is the canonical map in
(\ref{CanonicalBundleMap}).

In our applications, $\tilde V$ will have some additional
properties. We will be interested in the Fourier-Mukai transform of
a bundle $V$ with $c_1(V)=0$. The restriction to the generic fiber
$E$ should be semi-stable and degree zero. We define
\be \tilde V \ \equiv \ V \otimes \cO(\sigma_{B_2}) \ee
Given an elliptic fiber $E$, a generic $V$ splits as a sum of $r =
{\rm rank}(V)$ degree zero line bundles on $E$:
\be V_E \ \simeq\ \cO_E(p_1 - p_\infty) \oplus \ldots \oplus
\cO_E(p_r - p_\infty) \ee
The points $\{p_1, \ldots, p_r\}$, when varied over the base, sweep
out the spectral cover of $V$, and $p_\infty$ is the point that lies
on the zero section. (We denoted it by $p_\infty$ in order to
clearly distinguish it from the $p_i$). Then we have
\be \tilde V_E \ \simeq \ \cO_E(p_1) \oplus \ldots \oplus \cO_E(p_r)
\ee
and
\be H^0(E,\tilde V_E)\ =\ (s_1, \ldots ,s_r) \qquad H^1(E,\tilde
V_E)\ =\ 0 \ee
where $s_i$ is the unique section of $\cO_E(p_i)$, which vanishes at
$p_i$.

Let us examine the map (\ref{CanonicalBundleMap}) on the fibers over
a point $p \in Z$, and define $E_p = \pi^{-1}\pi(p)$. The fiber of
$\pi^*\pi_*\tilde V$ is
\be \pi^*\pi_*\tilde V_p \ = \ H^0(E_p, \tilde V_{E_p}) \ee
This is spanned by the $r$ sections $(s_1, \ldots s_r)$, and the map
(\ref{CanonicalBundleMap}) evaluates them at $p$. The sections are
linearly independent away from $\{p_1, \ldots, p_r\}$, so we can
represent $\Psi$ as
\be \Psi_E\ \sim \ \left(
  \begin{array}{ccc}
    s_1(p) & 0 & 0 \\
    0 & \ddots & 0 \\
    0 & 0 & s_r(p) \\
  \end{array}
\right) :\ \cO_E^r \ \to \ \tilde V_E \ee
The spectral cover is given by the zero locus of the $(s_1, \ldots
s_r)$, i.e. it is identified with the locus
\be \det(\Psi) \ = \ 0 \ee

All this is of course completely analogous to the map $\lambda I -
\Phi$ that we encountered for Higgs bundles; after diagonalizing
$\lambda I - \Phi$ as in equation (2.7) of part I, 
the diagonal entries behave like the sections $s_i$. Indeed, Higgs
bundles are not necessarily valued in a line bundle, but can take
values in more general objects, like an elliptic curve. This was
one of the main ideas in the adaptation of spectral cover methods
for Higgs bundles to the heterotic string
\cite{Friedman:1997yq,DonagiCovers}.

In particular, if the point $p$ does not lie on the spectral cover,
then (\ref{CanonicalBundleMap}) yields an isomorphism on the fibers.
But if the point $p$ does lie on the spectral cover, then the map
(\ref{CanonicalBundleMap}) has rank $r-1$. As in equation
(2.6) of part I 
for conventional Higgs bundles, it is natural to
define a sheaf ${\cal L}$ as the
cokernel of $\Psi$:
\be\label{LCokerDef} 0\ \ \to\  \ \pi^*\pi_*\tilde V\ \ \to\ \
\tilde V \ \ \to\ \ {\cal L}\ \ \to\ \ 0\ee
By the observations above, ${\cal L}$ is a rank one sheaf supported
on the spectral cover. Given the generality of the construction, it
must be essentially equal to the Fourier-Mukai transform of $V$.
(After submission, we noticed that this claim was also made in
\cite{FriedmanMorgan}).

Let us briefly review some generalities about the Fourier-Mukai transform.
A nice review is \cite{AndRuip}. We
define two projections, $p_{1,2}: Z\times_S Z \to Z$ on the first
and second factor respectively. We also define the Poincar\'e sheaf:
\be {\cal P}\ =\ \cO_{Z\times_B Z}(\Delta - Z \times \sigma_{B_2} -
\sigma_{B_2} \times Z) \otimes p_1^*\pi^*K_B^{-1}\ee
Then the Fourier-Mukai transform of a sheaf $V$ is defined to be
\be
 {\sf FM}^\bullet(V)\ \equiv\ R^\bullet p_{2*}(p_1^*V \otimes
{\cal P}) \ee
It is strictly speaking a complex, but if $V$ is reasonably well-behaved then
this complex is non-zero in only one degree, and we simply get a coherent sheaf.
We may also define the inverse transform
\be
\widehat{ {\sf FM}}^{i}(V)\ \equiv\ R^{i-1} p_{2*}(p_1^*V \otimes
{\cal P}') \ee
where
\be
{\cal P}' \ = \ {\cal P}^\vee \otimes p_1^*\pi^*K_{B_2}^{-1}
\ee
Since these definitions may look somewhat intimidating, let us explain some of the
intuition by restricting to a given elliptic fiber $E$.

The Poincar\'e sheaf ${\cal P}$ is the universal
line bundle on $E \times E^\vee$ such that the restriction to
$\sigma \in E^\vee$ gives the line bundle $\cO_E(\sigma - p_\infty)$
on $E$. This is the analogue of the factor $e^{ikx}$ in the ordinary
Fourier transform. Here $E^\vee$ is the dual elliptic
curve, which is isomorphic to $E$ itself. Therefore on each $E$, the
Fourier-Mukai transform is given by tensoring $V_E$ with a flat line
bundle $L_\sigma = \cO_E( p_\infty-\sigma)$, where $\sigma$ is a
coordinate on $E^\vee$, and then taking the cohomology $H^1(V_E
\otimes L_\sigma)$. In other words, as a line bundle it is defined
by assigning to a point $\sigma \in E$ the vector space
\be \sigma\ \to \ H^1(V_E \otimes L_\sigma) \ee
Now let us go back to the situation where the restriction decomposes as
$\tilde V_E \cong L_1 \oplus .. \oplus L_r$, where each $L_i$ is of
degree one. We have $L_i \simeq \cO(p_i - p_\infty)$, but we will
not specify the isomorphism explicitly because it depends on a
parameter which may vary over the base. Then we have
\be V_E\ \cong\ L_1(-p_\infty)\oplus \ldots L_r(-p_\infty) \ee
Since $h^1(L_i(-\sigma)) = h^0(L_i(-\sigma))^\vee = 1$ if $\sigma =
p_i$, and zero otherwise, we see that the Fourier-Mukai transform of
such a bundle $V$ is
supported on $\sigma = p_1, \ldots, p_r$, and the fibers of the dual sheaf
at these points are given
by $H^1(L_i(-p_i))$ respectively.

Now let us show we recover the same fiberwise structure from the
cokernel sequence (\ref{LCokerDef}).
The spectral cover
intersects $E$ in the points $p_1, \ldots,p_r$. Thus the fiber of
${\cal L}$ is given by the fiber of $L_i$ at $p_i$. This is clearly
isomorphic to $H^0(L_i|_{p_i})$. From the long exact sequence
associated to
\be 0 \ \ \to \ \ L_i \otimes \cO(-p_i)\ \ \to\ L_i\ \  \to\ \
L_i|_{p_i}\ \ \to \ \ 0 \ee
we see that $H^0(L_i|_{p_i})$ lifts to $H^1(L_i \otimes \cO(-p_i))$.
This is the same as what we got above from ${\sf FM}^1(V)|_E$.

Thus if $V$ is given by a holomorphic bundle on $Z$ which is semi-stable
on generic fibers, we expect that ${\sf FM}^1(V)$ and ${\cal L}$
must agree up to tensoring with a line bundle pulled-back from the
base. We will find the appropriate `normalization' below. We have
not proven this conjecture but one can do additional consistency checks.
One such check will be
done in section \ref{JordanBlockSpectralData}.

One way to find the `normalization' is as follows. With the
definition of ${\cal P}$ as above, we have ${\sf FM}^1(\cO_Z) =
\sigma_{B_2*}K_{B_2}$ \cite{AndRuip}. From (\ref{LCokerDef}) with $V = \cO_Z$ we
get ${\cal L} = \sigma_{B_2*}N_{B_2}$, i.e. the normal bundle on
$B_2$. Since $Z$ is a Calabi-Yau three-fold, these are the same. So
we have
\be {\sf FM}^1(V)\ =\ {\cal L} \ee
In some sense it would be more natural to twist the Poincar\'e sheaf
by $\pi^*K_B^{-1}$ so that ${\sf FM}^1(\cO_Z)= \cO_{B_2}$, but at
any rate this is a matter of convention.

Let us note two important and useful properties of ${\sf FM}$.
The first is the Parseval
formula
\be \Ext_X^i(F,G) \ = \ \Ext_{\hat X}^i({\sf FM}(F),{\sf FM}(G)) \ee
The second is that given a short exact sequence of sheaves
\be 0 \ \to \ F \ \to G \ \to K \ \to 0 \ee
we obtain the long exact
sequence:
\be \to \ {\sf FM}^{i-1}(K) \ \to \ {\sf FM}^i(F)\ \to \ {\sf
FM}^i(G)\ \to \ {\sf FM}^i(K)\ \to \ {\sf FM}^{i+1}(F)\ \to  \ee
These properties will be useful in the following.

The conclusion of our discussion is that we can find the spectral sheaf
if we can explicitly write the canonical evaluation map
(\ref{CanonicalBundleMap}). So far our discussion was general. Now
we will specialize to the case of monads.

Reference \cite{Bershadsky:1997zv} provided an algorithm for writing
down the spectral cover associated to a monad. This was based on the
work of \cite{Friedman:1997ih}, which showed that the spectral cover
corresponds to the zero locus of the determinant $ \det(\Psi) =  0$
of the map (\ref{CanonicalBundleMap}), as we also discussed above.
We now want to extend this to write down the spectral sheaf. The
algorithm of \cite{Bershadsky:1997zv} actually provides a set of
sections which generate $(s_1, \ldots ,s_r)$. By our previous
discussion and making some adaptions of \cite{Bershadsky:1997zv} we
then also recover the spectral sheaf, constructed as the cokernel.
We discuss this more explicitly in subsection \ref{ThreefoldMonad}.

\newsubsection{Jordan blocks}

\subseclabel{JordanBlockSpectralData}

In order for $V$ to admit a spectral cover description, its
restriction to the generic fiber $E$ should be semi-stable and
degree zero. If the restriction $V|_E$ to an elliptic fiber $E$ were
unstable, then the Fourier-Mukai transform of $V$ is supported on
the whole fiber, so this should not happen generically. So far we
seem to have assumed that any degree zero semi-stable bundle would
decompose as a direct sum of line bundles:
\be V_E \ \simeq \ \cO(p_1 - p_\infty) \oplus \ldots \oplus \cO(p_r
- p_\infty) \ee
Although this is generically the correct situation, it is not the
most general possibility. There also exist semi-stable degree zero
bundles on $E$ that are not decomposable as a sum of line bundles.
In fact, it will soon be clear that this situation occurs in
codimension one on the base even for generic spectral covers, namely
at the branch locus of the cover. In particular, the restriction of
the hermitian Yang-Mills connection on $V$ fails to be flat on these
fibers.

The general classification of semi-stable bundles on $T^2$ is due to
Atiyah \cite{AtiyahElliptic}. For each integer $r$ there exist rank
$r$ bundles on $E$ that are not decomposable. Consider the following
extension sequence on an elliptic curve:
\be 0\ \to\ \cO_E\ \to\ F_2\ \to\ \cO_E\ \to\ 0 \ee
Since $\Ext^1(\cO,\cO) = H^1(\cO) = {\bf C}$, we have two
possibilities: either $F_2 \sim \cO_E \oplus \cO_E$, or $F_2$ is the
unique non-trivial rank two extension. Similarly we can consider the
unique non-trivial extensions on $T^2$:
\be 0\ \to\ F_{r-1}\ \to\ F_r\ \to\ \cO_E\ \to\ 0 \ee
The most general semi-stable bundle of slope zero is a sum of
factors of the form $F_r \otimes L$, where $L$ is a degree zero line
bundle and we took $F_1 = \cO_E$.

Since the spectral cover apparently exists for such more general
bundles, let us try to see what it looks like. The bundle $F_2$ is
an extension of $\cO_E$ by itself, so the Fourier-Mukai transform of
$F_2$ will have the {\it same} support as $\cO_E \oplus \cO_E$. A
similar statement obviously holds for $F_r$ with $r>2$. Thus the
question arises how the spectral cover description distinguishes
between $F_2$ and $\cO_E \oplus \cO_E$. This has been previously
explained in \cite{Aspinwall:1998he} (or even earlier in the math
literature), and with our preparation in the previous sections it
should not be hard to guess: the Fourier-Mukai transform of $F_2$
has a nilpotent Higgs VEV, and the Fourier-Mukai transform of $\cO_E
\oplus \cO_E$ has vanishing Higgs VEV.

Let us check this more explicitly. Chasing through the definition of
the Fourier-Mukai transform, we see that the fibers of the spectral
sheaf over a point $\lambda \in E$ are found as follows: we tensor
$F$ with a flat line bundle $\cO(\lambda - p_\infty)$, and then take
global sections. Clearly this does not yield $\cO_{p_\infty} \oplus
\cO_{p_\infty}$ but the non-trivial extension of $\cO_{p_\infty}$ by
itself. We have seen this before in section 2.3 of part I:
this corresponds to the structure sheaf of a fat point, or
equivalently to a nilpotent Higgs VEV.

In the remainder of this
subsection, we will show that the construction of the Fourier-Mukai transform
as coker$(\Psi)$ also reproduces this. The discussion is parallel
with the discussion of section 2 of part I. For convenience we
consider again the rank two case. We consider the twisted bundle
\be \tilde V_E \ = \  F_2 \otimes \cO(p_\infty) \ee
Now note that $\tilde V_E$ is the extension of $\cO(p_\infty)$ by
itself:
\be\label{Rank2Slope1Extension} 0 \ \to \ \cO(p_\infty) \ \to \ F_2
\otimes \cO(p_\infty) \ \to\ \cO(p_\infty) \ \to \ 0 \ee
In particular, from the long exact sequence we see that
\be H^0(\tilde V_E)\ =\ 2, \qquad H^1(\tilde V_E)\ =\ 0 \ee
Thus again $H^0(\tilde V_E) $ is generated by two sections, just as
in the decomposable case. More generally when $\tilde V_E = F_r
\otimes \cO(p)$, we have $H^0(\tilde V_E)=r $ and $H^1(\tilde V_E)=0
$.

Just as before, we use the sections of $H^0(\tilde V_E) $ to define
a map
\be \Psi_E: \, \cO_E \oplus \cO_E \ \to\ \tilde V_E \ee
and the spectral sheaf will be the cokernel of this map. We pick a
local coordinate $\lambda$ on $E$ such that $\lambda = 0$
corresponds to $p_\infty$. The two sections are linearly independent
away from $p_\infty$, so the spectral sheaf will be localized at
$\lambda=0$, and to figure out the precise description we only need
the form of the map near $\lambda=0$.

Let us first consider the sections of $F_2$. From the exact sequence
\be 0\ \to\ \cO\ \to F_2\ \to\ \cO\ \to\ 0 \ee
we get
\be
 0 \ \to \ H^0(\cO_E) \to H^0( F_2) \to H^0( \cO_E) \to H^1( \cO_E)
\ee
The last map is the extension class, which is by definition
non-zero, so we see that $H^0(F_2)$ is one-dimensional. It has a
unique section, obtained from $\cO_E$ by injecting it into $F_2$.

Now we come to the sections of $\tilde V_E$. The first section of
$\tilde V_E$ is inherited from $F_2$, i.e. we take the unique
section of $F_2$ and tensor it with a section of $\cO(p_\infty)$.
The unique section of $\cO_E$ is locally just given by $1$, so up to
a change of basis, locally we can always represent the section of
$F_2$ as $(1,0)$. Furthermore the section of $\cO(p_\infty)$ can be
represent as $\lambda$, so we can represent the first section of
$\tilde V_E$ as
\be s_1\ =\ \lambda \cdot (1,0)\ =\ (\lambda,0) \ee
The map from $F_2$ to $\cO_E$ is given by projection on the second
argument.

The second section of $\tilde V_E$ maps to a section of the quotient
in (\ref{Rank2Slope1Extension}). In this case, the quotient is also
$\cO(p_\infty)$, and its section is given by $\lambda$, so the
second section of $\tilde V_E$ takes the form
\be s_2\ =\ (*,\lambda) \ee
We further know that this section cannot be inherited from $F_2$, so
it is not proportional to $\lambda$ (in particular it can not vanish
at $\lambda=0$). This only leaves the following possibility:
\be s_2\ =\ (1,\lambda) \ee
We conclude that near $\lambda=0$, we can represent $\Psi_E$ as
\be \Psi_E\ =\ \left(
  \begin{array}{cc}
    \lambda & 1 \\
    0 & \lambda \\
  \end{array}
\right) \ee
The spectral sheaf, restricted to $E$, is the cokernel of $\Psi_E$.
As we have seen before, this is precisely the structure sheaf
$\cO_{2p_\infty}$.

Clearly we can generalize this to the higher rank versions. We have:
\be H^0(E,F_k\otimes \cO(p_\infty)) = \vev{s_1, .., s_k}, \qquad
H^1(E,F_k\otimes \cO(p_\infty))= 0. \ee
and the map $\Psi_E$ near $\lambda = 0$ consists of a rank $k$
Jordan block:
\be \Psi_E\ \simeq\ \left(
  \begin{array}{ccccc}
    \lambda & 1 &  &  & \\
    0 & \lambda &  & & \\
    &  &  \ddots     &   &    \\
    & &  & \lambda & 1  \\
    & &  & 0 & \lambda  \\
  \end{array}
\right) \ee
The cokernel of $\Psi_E$ is the structure sheaf of a fat point of
length $k$, $\cO_{k\,p_\infty}$.

\newsubsection{Review of linear sigma models}

\subseclabel{LSMReview}

Linear sigma models are one of the prime methods for constructing
exactly conformal $(0,2)$ CFTs, which can be used to build vacua for
the heterotic string. These models exhibit many interesting
properties. In geometric phases they correspond to bundles defined
by a monad. A residue theorem of \cite{Beasley:2003fx} shows that
conformal invariance is not spoiled by world-sheet instantons. Let
us briefly recall some basic aspects of such constructions. We refer
to \cite{Witten:1993yc} for details.

We consider a two-dimensional $U(1)$ gauge theory with $(0,2)$
supersymmetry. The right-moving fermionic coordinate is denoted by
$\theta$. The main multiplets are $(0,2)$ chiral fields, defined by
the condition
\be \overline{\cal D}_+ \Phi \ = \ 0 \ee
and $(0,2)$ Fermi multiplets, defined by the condition
\be \overline{\cal D}_+ \Lambda \ = \ E \ee
where
\be \overline{\cal D}_+ \ = \ -{\del\over \del \overline{ \theta}} +
i \,\theta\, {\cal D}_{\bar z} \ee
is the spinorial covariant derivative, and $E$ is some (composite)
chiral field. The $U(1)$ charges of chiral fields are denoted by
$q$, and charges of fermionic multiplets are denoted by $\tilde q$.
Finally we have the $(0,2)$ vector multiplet, whose field strength
will be denoted by $\Upsilon$.

The matter fields of our model are given by $(0,2)$ chiral
superfields $\Phi = \{ X_i, \ P, \ \Sigma\}$ with $U(1)$ charges
$q_i>0$, $q_P < 0$, and $q_\Sigma = 0$ respectively. We further have
$(0,2)$ Fermi fields $\Lambda = \{ \Lambda_a, \ \Gamma \}$ with
charges $\tilde q_a > 0$ and $\tilde q_\Gamma < 0$. These charges
are subject to the relations
\be\label{Mon1Constraints} \tilde q_\Gamma\ =\ -\sum q_i, \qquad
q_P\ =\ -\sum \tilde q_a. \ee
The action is of the schematic form
\be\label{LSMAction} \int d^2 z d^2\theta  \left[{1\over 8e^2} \bar
\Upsilon \Upsilon + \bar \Phi \bar {\cal D}_{ z} \Phi + \bar \Lambda
\Lambda\right] + {t\over 4} \int d^2 z d\theta\, \Upsilon|_{\bar
\theta = 0} + \int d^2 z d\theta\, W(\Phi,\Lambda)|_{\bar \theta =
0}\ +\ h.c. \ee
where we used $\Phi$ and $\Lambda$ to denote general chiral and
Fermi fields, respectively. Here the parameter
\be t \ =\ i\,r + {\vartheta\over 2\pi} \ee
contains the Fayet-Iliopoulos parameter and theta-angle. The
integral over half of superspace is invariant if the integrand is
chiral, i.e. annihilated by $\overline{\cal D}_+ $. We take the
superpotential of the form
\be  W(\Phi,\Lambda)\ = \   \Gamma\, G(X_i) + \Lambda_a P J^a(X_i)
\ee
In order for this to be gauge invariant, $G$ is of degree $-\tilde
q_\gamma$ in the $X_i$. Similarly $J^a$ is also chiral and of degree
$-q_p - \tilde q_a$ in the $X_i$, and in order to get a chiral
integrand the $J^a$ are constrained by a relation that we mention
momentarily. We further take
\be \overline{\cal D}_+ \Gamma\ =\ \Sigma\, P\, E_\Gamma(X_i),
\qquad \overline{\cal D}_+ \Lambda_a\ =\ \Sigma\, E_a(X_i) \ee
Then the requirement $\overline{\cal D}_+  W = 0$ implies that the
$E$'s and $J$'s are subject to
\be E_a J^a \ =\ -E_\gamma\, G \ee
The action (\ref{LSMAction}) leads to a $D$-term potential of the
form
\be V_D \simeq {e^2\over 2} \left(\sum q_i|x_i|^2 + \tilde
q_\Gamma|p|^2 - r\right) + |G|^2 + |p|^2\sum |J_a|^2 +  |\sigma|^2
\left(\sum |E_a|^2 + |p E_\Gamma|^2\right) \ee
In the geometric regime $r
>>0$, this flows to a non-linear sigma-model on the Calabi-Yau
hypersurface $G(x_i)=0$ in ${\bf WP}_{q_1, \ldots, q_n}$, and the
surviving massless right-moving fermions $\psi_{X_i}$ take value in
the tangent bundle of the Calabi-Yau. We further have additional
left-moving fermions $\lambda_a$ together with some mass terms
\be \psi_P\,\lambda_a J^a(x)\ +\ \overline{\psi}_\Sigma\, \lambda_a
{ \overline{E}_a} \ee
induced from the Yukawa couplings. The constraint coming from the
second mass term can be formulated holomorphically by introducing
fermionic gauge equivalences:
\be \lambda_a\ \sim\ \lambda_a + E_a s. \ee
Then the surviving massless left-moving fermions live in a bundle
$V$ which is given by the cohomology of the monad
\be\label{LSMMonad}
 0\ \ \to\ \ \cO\ \ \mathop{\longrightarrow}^{E_a}\ \ \bigoplus \cO(\tilde q_a)\
\ \mathop{\longrightarrow}^{J^a}\ \
 \cO(-q_P)\ \ \to\ \ 0
\ee
This is a complex on the Calabi-Yau hypersurface $G=0$, because
$E\cdot J = 0 \ {\rm mod}\, G$. In the special case where
\be
 E_i\ =\ q_i X_i, \qquad E_\Gamma\ =\ -\tilde q_\Gamma, \qquad J^i \ =\ {\del G \over \del X_i}
\ee
this is the Euler sequence for the tangent bundle, and yields a
left-right symmetric model. The above considerations may be easily
generalized to multiple $U(1)$'s (allowing for multi-degrees),
multiple $\Gamma$'s (allowing for complete intersections), and
multiple $\Sigma$'s (allowing for extra fermionic gauge
equivalences).

Not any monad defines a linear sigma model. We have already seen the
conditions (\ref{Mon1Constraints}), which correspond to absence of
anomalies for the right-moving $U(1)_R$ symmetry and an additional
left-moving global $U(1)$ symmetry. Although these are global
symmetries, we need them to be non-anomalous in order to construct a
string vacuum. Geometrically they can be interpreted as $c_1(T) =
c_1(V) = 0$. To cancel the $U(1)$ gauge anomaly, we get the further
constraint $\tilde q^2 - q^2 = 0$, or more explicitly:
\be \sum \tilde q_i \tilde q_i  + \tilde q_\gamma \tilde q_\gamma -
\sum q_i q_i - q_p q_p \ = \ 0 \ee
This is closely related, although in general not quite equivalent to
the condition $c_2(T) = c_2(V)$.

\newsubsection{Monads over an elliptic three-fold}

\subseclabel{ThreefoldMonad}

In this section we would like to explain the algorithm for finding
the spectral sheaf of a monad, generalizing the algorithm for
finding the spectral cover in \cite{Bershadsky:1997zv}. We assume of
course that the monad bundle is semi-stable on generic elliptic
fibers, because otherwise the Fourier-Mukai transform is not
supported on a divisor. This is also reasonable because generic
bundles with vanishing first Chern class will have this property.

\newsubsubsection{Summary of the algorithm}

Let us state the strategy at the outset. We
consider general complexes of the form
\be
 0\ \to\ \cO_Z^{\oplus q}\ \ \mathop{\longrightarrow}^E\ \ {\cal H} \ \mathop{\longrightarrow}^J\  {\cal N}\ \to\ 0
\ee
The strategy will be to first study the bundle ${\cal K}$ given by
the kernel of the map $J$, without modding out by $E$. That is,
${\cal K}$ is defined by the short exact sequence
\be
 0\ \to\ {\cal K}\ \to \ {\cal H} \ \mathop{\longrightarrow}^J\  {\cal N}\ \to\ 0
\ee
Working with ${\cal K}$ illustrates most of the important points,
and adapting it for $V= ker(J)/im(E)$ is only a small modification
of the procedure. Thus we will split the algorithm into two steps:
\begin{description}
  \item[{{\it Step i.}}] The bundle ${\cal K}$ is an extension of
$V$ by $\cO_Z^{\oplus q}$, where $V$ is the bundle we are eventually
interested in. It is also semi-stable and degree zero on the
elliptic fibers if $V$ is. Thus we can ask for the Fourier-Mukai
transform of ${\cal K}$. We claim that the spectral sheaf of ${\cal
K}$ is given by the cohomology ${\rm ker}(J)/{\rm im}(A)$ of the
following complex:
\be 0 \ \ \to\ \ \pi^*\pi_*\tilde {{\cal K}}\ \
\mathop{\longrightarrow}^A \ \ \widetilde {\cal H}\ \
\mathop{\longrightarrow}^J\ \ \widetilde{\cal N}\ \ \to\ \ 0 \ee
In particular, the support of ${\sf FM}^1({\cal K})$ can be
recovered from the determinant of this complex. The map $A$ is the
composition of the canonical map $\pi^*\pi_*\tilde {\cal K} \to
\tilde{\cal K}$ followed by the canonical inclusion $\tilde{\cal K}
\to \tilde{\cal H}$.
  \item[{{\it Step ii.}}] Once we have ${\sf
FM}^1({\cal K})$, we can recover ${\sf FM}^1(V)$ from the short
exact sequence
\be 0\ \to\ {\sf FM}^1(\cO_Z)^{\oplus q} \ \to\ \  {\sf FM}^1({\cal
K})\ \ \to\ \ {\sf FM}^1(V)\ \  \to\ \ 0 \ee
In particular, the support of ${\sf FM}^1({\cal K})$ is simply the
union of the support of ${\sf FM}^1(\cO_Z)^{\oplus q}$, which
consists of $q$ copies of the zero section $\sigma_{B_2}$ of $Z$,
and the support of ${\sf FM}^1(V)$.
\end{description}
As before, recall that $\tilde V \equiv V \otimes \cO(\sigma)$ for any bundle $V$.
Below we will explain these steps in more detail. We will illustrate
the procedure with a concrete class of examples.

\newsubsubsection{Derivation and example}

For the class of examples we take some models that were considered
in \cite{Bershadsky:1997zs}. The Calabi-Yau is an elliptic fibration
over the Hirzebruch surface ${\bf F}_n$, with $n=0,1,2$ so that $Z$
is smooth. This can be embedded as a Weierstrass model in a ${\bf
WP}_{1,2,3}$ fibration over ${\bf F}_n$, so the linear sigma model
will have three $U(1)$ gauge fields. We will denote the homogeneous
coordinates collectively by $x_i$, $i = 1, \ldots, 7$, the
homogeneous coordinates for ${\bf F}_n$ by $\{u_0,u_1;v_0,v_1 \}$,
and the homogenous coordinates for ${\bf WP}_{1,2,3}$ by
$\{x,y,z\}$. Then $Z$ is given by a hypersurface equation
\be G\ \equiv\ -y^2+ x^3 + f(u,v) x z^4 + g(u,v) z^6\ =\ 0 \ee
Over this Calabi-Yau we consider the following monad:
\be\label{V6Sequence}
 0\ \to\ \cO_Z\ \ \mathop{\longrightarrow}^E\ \ {\cal H} \ \mathop{\longrightarrow}^J\  {\cal N}\ \to\ 0
\ee
where we put
\be {\cal H} \ = \ \sum_i \cO_Z(\sum_J n_{iJ} D_J), \qquad {\cal N}
\ = \ \cO_Z(\sum_J m_J D_J)  \ee
and $m_J = \sum_i n_{iJ}$. This is precisely the type of monad one
gets from gauged linear sigma models, cf. equation (\ref{LSMMonad}).
We will take our bundle $V_5$ to be a deformation of $TZ \oplus
\cO_Z \oplus \cO_Z$. Then the number of left-moving fermionic fields
$\Lambda_a$, which describe the bundle $V_5$, is the same as the
number of bosonic fields $X_i$ for the underlying toric manifold,
and their $U(1)^3$ charges are the same also:
\be n_{iJ} \ = \ \left(
  \begin{array}{ccccccc}
    1 & 1 & n & 0 & 4+2n & 6+3n & 0 \\
    0 & 0 & 1 & 1 & 4 & 6 & 0 \\
    0 & 0 & 0 & 0 & 2& 3 & 1 \\
  \end{array}
\right) \ee
The divisors $D_J$ are given by
\be D_1 = \{u_1=0\}, \quad D_2 = \{v_1 = 0\}, \quad D_3 = \{z=0 \}
\ee
Let us denote the base and fiber of the Hirzebruch by $b$ and $f$,
with $b^2 = -n$, $f^2 = 0$, $b\cdot f = 1$.  Then $D_1$ corresponds
to $\pi^*f$ and $D_2$ corresponds to $\pi^*b$. The divisor $D_3$
corresponds to the zero section $\sigma_{B_2}$ of the elliptic
fibration.

Let us first consider the number of generations in such a model.
When $J$ is given by the partial derivatives
\be  J\ =\ (\del G/\del \vec{x})^T \ee
then the bundle $V$ is a non-trivial extension of $TZ$ by $\cO_Z
\oplus \cO_Z$. Since the Euler character is additive in an exact
sequence, we have
\be \chi(V) \ = \ \chi(TZ) + 2 \chi(\cO_Z) \ = \ \chi(TZ) \ee
Further, the net number of generations does not change under
continuous deformations. Thus the net number of generations is given
by the Euler characteristic of $Z$. Assuming that $Z$ is a smooth
Weierstrass model with one section, we have \cite{Andreas:1998zf}
\be \chi(Z) \ = \ -60\, c_1(S)^2 \ee
Since $c_1^2=8$ for all the Hirzebruch surfaces, we get $480$
generations. Needless to say this is far from realistic, but our
point here is only to illustrate the technology.

In our case, ${\cal H}$ has rank seven and ${\cal N}$ has rank one,
so the kernel of $J$ has rank six and will be denoted by ${\cal
K}_6$. The bundle ${\cal K}_6$ is defined by the short exact
sequence:
\be\label{K6Sequence}
 0\ \to\ {\cal K}_6\ \to \ {\cal H} \ \mathop{\longrightarrow}^J\  {\cal N}\ \to\ 0
\ee
Now one may try to simplify the maps by using the freedom to make
field redefinitions. Apart from coordinate redefinitions of the
variety, we have additional field redefinitions of the form
\be \Lambda_a \ \to \ \Lambda_a + \sum p_{ab}(X_i) \Lambda_b \ee
where $p_{ab}$ is a matrix of polynomials of the appropriate degrees
in the chiral fields $X_i$, whose bosonic components yield the
coordinates $x_i$ on the ambient variety. These field redefinitions
correspond to bundle automorphisms $\Ext^0({\cal H},{\cal H})$, i.e.
symmetries of the rank $7$ bundle ${\cal H}$. Under such
automorphisms, we have
\be J^a \ \to \ J^a + p_{ba}(x_i) J^b \ee
According to \cite{Bershadsky:1997zs}, generically we can set the
first four entries of $J$ equal to zero by field redefinitions. We
were not able to reproduce this, but for illustrative purposes we
take $J$ to be of the following form:
\be
\begin{array}{ccccccccl}
  J & = & \left(\, 0,\right. & 0, & 0, & 0, & {\del G\over \del x},  & {\del G\over \del y}, & \left.
  {\del G\over \del z}+ P_{8,8+4n} xz^3
  \,\right)
\end{array}
\ee
where $P_{8,8+4n}$ is a polynomial of bidegree $(8,8+4n)$ on ${\bf
F}_n$. This family contains the essential features and keep the
calculations simple enough to write out here.

As in section \ref{BundleGeneralities}, we now twist the sequence
(\ref{K6Sequence}) by $\cO(D_3)$. As discussed, the Fourier-Mukai
transform of ${\cal K}_6$ is given by the short exact sequence
\be 0 \ \ \to\ \ \pi^*\pi_*\tilde {\cal K}_6\ \
\mathop{\longrightarrow}^{\Psi_6}\ \ \tilde {\cal K}_6\ \  \to\ \
{\cal L}_6\ \ \to\ \ 0 \ee
We have realized $\tilde {\cal K}_6$ as the kernel of $J$, but we do
not yet have $\pi^*\pi_*\tilde {\cal K}_6$ or an explicit
representative for $\Psi_6$. To get $\pi^*\pi_*\tilde {\cal K}_6$,
we take cohomology along the elliptic fiber, that is we use (\ref{K6Sequence})
(twisted by $\cO(D_3$)) to get the long exact sequence
\be\label{kerJ*def}  0 \ \ \to\ \ \pi^*\pi_*\tilde {\cal K}_6\  \ \mathop{\to}^i \ \
\pi^*\pi_*\tilde {\cal H}\ \ \mathop{\longrightarrow}^{J_*}\ \
\pi^*\pi_*\widetilde {\cal N}\ \ \to\ \ 0 \ee
where as usual, a tilde denotes twisting by $\cO(D_3)$:
\be \tilde{\cal H} \ = \ {\cal H} \otimes \cO(D_3) \ = \ \sum_i
\cO(\sum_J n'_{iJ} D_J), \qquad \tilde {\cal N} \ = \ \cO(\sum_J
m'_J D_J) \ee
and
\be n'_{iJ}\ = \ n_{iJ} + \delta_{J3}, \qquad m'_J\ = \ m_J +\,
\delta_{J3} \ee
Our long exact sequence truncates after just three terms because
$H^1(\tilde {\cal K}_6|_{T^2})$ vanishes for stable bundles of
positive degree on the elliptic fiber, as explained in section
\ref{BundleGeneralities}. Thus we have realized $\pi^*\pi_*\tilde
{\cal K}_6$ as the kernel of $J_*$. Note that the kernel of $J_*$ is
$13-7=6$ dimensional, the same as the rank of ${\cal K}_6$ and
$\tilde {\cal K}_6$, as expected by the general discussion in
section \ref{BundleGeneralities}.

Finally then, we need the map $\Psi_6$. Recall that this is the
canonical evaluation map $\pi^*\pi_*\tilde {\cal K}_6 \to \tilde
{\cal K}_6$. Now we have realized $\pi^*\pi_*\tilde {\cal K}_6$ as
the kernel of $J_*$ in $\pi^*\pi_*\tilde{\cal H}$ and we have
realized $\tilde {\cal K}_6$ as the kernel of $J$ in $\tilde{\cal
H}$, so what we need is the evaluation map from
$\pi^*\pi_*\tilde{\cal H}$ to $\tilde{\cal H}$.

To do this, we need to know the restriction of $\cO(D_J)$ to $T^2$.
$D_3$ intersects every $T^2$ at a single distinguished point at
infinity on the elliptic fiber, so we define $\cO(D_3)|{T^2} =
\cO(1)$. $D_1$ and $D_2$ are disjoint from $T^2$ for almost all
elliptic fibers, so at least on a dense set on the base we have that
$\cO(D_1)|_{T^2}$ and $\cO(D_2)|_{T^2}$ are the trivial line bundle
on $T^2$ (denoted by $\cO(0)$).

It is not hard to derive the following identifications:
\be
\begin{array}{rcl}
  \pi_*\cO_Z(a,b,0) & = & \cO_S(a,b) \\[2mm]
  \pi_*\cO_Z(a,b,k) & = & \cO_S(a,b) \otimes (\cO_S \oplus  K_S^{2}\oplus  K_S^{3}
  \oplus \ldots \oplus  K_S^{k}) \\
\end{array}
\ee
where the second line is for $k \geq 1$. We further have the
canonical map
\be \pi^*\pi_*\cO_Z(a,b,k)\ \to\ \cO_Z(a,b,k) \ee
which over an open subset is given by mapping local sections as
\be {\rm ev}:\ (p_0, p_2, p_3, \ldots p_k) \ \to \ p_0z^k + p_2
xz^{k-2} + \ldots + p_k x^{(k-1)/2}y \ee
for $k$ odd, and the last term given by $p_k x^{k/2}$ if $k$ is
even. Note that some of the $p_k$ may not extend to global sections,
which is why we have to work over an open subset. Equivalently, we
can work with meromorphic sections.

Putting it all together, we find that the spectral sheaf ${\cal
L}_6$ is given by the cohomology of the following complex:
\be 0 \ \ \to\ \ \pi^*\pi_*\tilde {\cal K}_6\ \
\mathop{\longrightarrow}^A \ \ \widetilde {\cal H}\ \
\mathop{\longrightarrow}^J\ \ \widetilde{\cal N}\ \ \to\ \ 0 \ee
Here $A$ is given by the composition $A = {\rm ev}_{\cal H} \circ i$
where $i$ is the inclusion in (\ref{kerJ*def}) and $ {\rm ev}_{\cal H}$ is the
canonical evaluation map $\pi^*\pi_*\widetilde{\cal H} \to
\widetilde{\cal H}$. Thus to find ${\cal L}_6$, we see that
everything eventually boils down to finding an explicit
representative for $A$. It is induced from the evaluation map ${\rm
ev}_{\widetilde{\cal H}}$, by restricting to the kernel of $J_*$.
Moreover we can calculate ${\rm ev}_{\widetilde{\cal H}}$ very
explicitly because ${\widetilde{\cal H}}$ is just a sum of line
bundles. This is a rather general result, and doesn't depend on the
specific class of examples we have chosen.

In the case of our examples, we can choose local bases for
$\pi^*\pi_*\tilde V_6$ and $ \widetilde {\cal H}$ so that the map
$A$ is represented as follows:
\be A =\left(
  \begin{array}{cccccc}
    z & 0 & 0 & 0 & 0 & 0 \\
    0 & z & 0 & 0 & 0 & 0 \\
    0 & 0 & z & 0 & 0 & 0 \\
    0 & 0 & 0 & z & 0 & 0 \\
    0 & 0 & 0 & 0 & {\del G\over \del y} & Q_3 \\
    0 & 0 & 0 & 0 & -{\del G\over \del x} & Q_4 \\
    0 & 0 & 0 & 0 & 0 & Q_2 \\
  \end{array}
\right) \ee
with
\ba Q_3 \is -3gP_8(4f-P_8) z^3 + (108 g^2 + f P_8^2) xz \eol Q_4 \is
{3\over 2} (108 g^2 + f P_8^2) yz  \eol
 Q_2 \is  2(27 g^2 + f^2 P_8) z^2 + 9 g (4f-P_8) x.
\ea
The matrix is $7\times 6$ as the source is $\pi^*\pi_*\tilde V_6$
which has rank $6$, and the target has rank $7$.

The first four columns are very easy to understand. Clearly each of
the first four  line bundle of the form $\pi^*\pi_*\cO(a,b,1)\cong
\pi^*\cO(a,b)$ in $\widetilde {\cal H}$ are in the kernel of $J_*$.
We have
\be \Hom(\pi^*\cO(a,b), \cO(a,b,1)) \ =\ H^0(\cO(0,0,1)) \ee
which is one dimensional and has a unique global section up to
rescaling, given by $z$. So we can choose a basis so that the first
four columns are as above.

The remaining two columns can not be written globally. We can write
them over open subsets $U$ of some covering of $Z$. We take $U_i$ to
be the complement of $k_i(u,v)=0$, where $k_i$ is a section of
$K_B^{-2}$. We can cover $Z$ by three such open sets.

As explained above, the spectral sheaf ${\cal L}_6$ is recovered as
${\rm ker}(J)/{\rm Im}(A)$. In particular, its support is given by
$\det(\Psi_6)=0$ where
\be \vec{M}_A \ = \ \det (\Psi_6)\, \vec{J} \ee
and $\vec{M}_A$ is the vector of minors of $A$. Therefore, the
equation of the spectral cover is given by
\be \det(\Psi_6)\ =\ z^4 Q_2. \ee
The fact that the support of ${\cal L}_6$ is reducible is hardly a
surprise. The first four line bundles of ${\cal H}$ are clearly in
the kernel of $J$, and their transform is a sum of four line bundles
supported on the zero section. The map $J$ only acts non-trivially
on the last three summands of ${\cal H}$, and its kernel on these
three summands transforms to a sheaf ${\cal L}_2$ supported on $Q_2
= 0$. The sheaf ${\cal L}_6$ is the sum of the four line bundles on
$z=0$ and ${\cal L}_2$.

Now we come to step two, where we want to also mod out by the image
of $E$ and recover $V_5$. That is, we now consider the short exact
sequence
\be 0\ \to\ \cO_Z\ \to\ \  {\cal K}_6\ \  \to\ \  V_5\ \  \to\ \  0
\ee
Applying the Fourier-Mukai transform, we get a long exact sequence,
which truncates to
\be 0\ \to\ {\sf FM}^1(\cO_Z)\ \to\ \  {\sf FM}^1(V_6)\ \ \to\ \
{\sf FM}^1(V_5)\ \  \to\ \ 0 \ee
We learn several things. First of all, the support of ${\sf
FM}^1({\cal K}_6)$ is the same as the union of the supports of ${\sf
FM}^1(\cO_Z)$ (which is just the zero section $\sigma_B$) and ${\sf
FM}^1(V_5)$. Thus to find the spectral cover for $V_5$, we simply
take the spectral cover we found for $V_6$ and divide by $z$ (the
equation for the zero section). In particular, the extra steps in
\cite{Bershadsky:1997zv} can be skipped.

To find the spectral sheaf rather than just the support, we also
need the map, which is given by ${\sf FM}^1(E)$. In practice
however, we are interested in the situation where $V_5$ is stable. A
necessary condition for stability is that $V_5$ has no sections. The
role of $E$ is to mod out by the sections of ${\rm ker}(J)$, i.e.
the $E$'s are simply given by all the generators of $H^0({\cal K}_6) =
\Hom(\cO_Z,{\cal K}_6)$. The dual statement is simply that ${\sf FM}^1(E)$
is given by all the generators of $\Hom({\sf FM}^1(\cO_Z), {\sf
FM}^1({\cal K}_6))$. So we do not have to translate $E$ explicitly through
the Fourier-Mukai transform.

In our toy example, $V_5$ is actually not stable, but things are
nevertheless very easy. The bundle ${\cal K}_6$ is the direct sum of
four degree zero line bundles pulled back from the base and a stable
rank two bundle $V_2$, whose spectral cover is given by $Q_2=0$.
Then $E$ is necessarily of the form
\be
\begin{array}{ccccccccl}
  E & = & \left(\,*, \right.& *, & *, & *, & 0, & 0, & \left. 0\,\right) \\[4mm]
\end{array}
\ee
The transform ${\sf FM}^1(\cO_Z)$ is supported on $z=0$ and so can
only map into the part of the spectral sheaf that is supported at
$z^4 = 0$. Furthermore, the $E$'s are then simply pull-backs of
sections of line bundles on the base, which we may also call $E$ (as
they are given by the same expressions). Therefore we get a short
exact sequence for a $U(3)$ bundle $U_3$ on $B_2$:
\be 0 \ \ \to \ \ \cO_{B_2}\ \ \mathop{\longrightarrow}^E \ \
\cO(1,0)_{B_2}^2 \oplus \cO(n,1)_{B_2} \oplus \cO(0,1)_{B_2} \ \ \to
\ \ {U_3} \ \ \to \ \ 0 \ee
The Fourier-Mukai transform of $V_5$ is therefore the sum of
$\sigma_{B_2*} U_3$ and ${\cal L}_2$.

\bigskip

\noindent {\it Acknowledgements:} We would like to thank L.~Anderson
for initial collaboration. MW would like to thank C.~Vafa and the
Heisenberg program of the German Science Foundation for financial
support during this project. RD acknowledges partial support by NSF grants 0908487 and
0636606. MW would also like to thank the Taiwan string
theory workshop, Bonn University and the KITPC in Beijing for the opportunity to present some of these
results.

\end{document}